\documentclass[english]{article}
\usepackage[utf8]{inputenc}
\usepackage[T1]{fontenc}
\usepackage{babel}
\usepackage{amsmath}
\usepackage{graphicx}
\usepackage{fancyhdr}
\usepackage{hyperref}
\hypersetup{
    colorlinks=true,
    linkcolor=red,
    filecolor=magenta,      
    urlcolor=blue,
}

\begin{document}

\title{\textbf{BOLD5000} \\ A public fMRI dataset of 5000 images}

\author{Nadine Chang\textsuperscript{1}, John A. Pyles\textsuperscript{1}, Abhinav Gupta\textsuperscript{1}, \\ Michael J. Tarr\textsuperscript{1}, Elissa M. Aminoff\textsuperscript{2{*}}}

\maketitle
\thispagestyle{fancy}

1. Carnegie Mellon University 2. Fordham University {*}corresponding author:
Elissa Aminoff (eaminoff@fordham.edu)

\begin{abstract}
% no longer than 170 words
% currently 176

Vision science, particularly machine vision, has been revolutionized by introducing large-scale image datasets and statistical learning approaches. Yet, human neuroimaging studies of visual perception still rely on small numbers of images (around 100) due to time-constrained experimental procedures. To apply statistical learning approaches that integrate neuroscience, the number of images used in neuroimaging must be significantly increased. We present BOLD5000, a human functional MRI (fMRI) study that includes almost 5,000 distinct images depicting real-world scenes. Beyond dramatically increasing image dataset size relative to prior fMRI studies, BOLD5000 also accounts for image diversity, overlapping with standard computer vision datasets by incorporating images from the Scene UNderstanding (SUN), Common Objects in Context (COCO), and ImageNet datasets. The scale and diversity of these image datasets, combined with a slow event-related fMRI design, enable fine-grained exploration into the neural representation of a wide range of visual features, categories, and semantics. Concurrently, BOLD5000 brings us closer to realizing Marr's dream of a singular vision science --- the intertwined study of biological and computer vision.

\end{abstract}

% (700 words maximum) 

\section*{Background \& Summary}

Both human and computer vision share the goal of analyzing visual inputs to accomplish high-level tasks such as object and scene recognition\cite{Yamins2016}. Dramatic advances in computer vision were driven over the past few years by massive-scale image datasets such as ImageNet \cite{Deng2009} and COCO \cite{Lin2014}. In contrast, the number of images used in most studies of biological vision remains extremely small. Under the view that future progress in vision science will rely on the integration of computational models and the neuroscience of vision, this difference in the number of stimuli used in each domain presents a challenge. How do we scale neural datasets to be commensurate with state-of-the-art computer vision, encompassing the wide range of visual inputs we encounter in our day-to-day lives?

The motivation to scale up neural datasets is to leverage and combine the wide variety of technological advances that have enabled significant, parallel progress in both biological and machine vision. Within biological vision, innovations in neuroimaging have advanced both the measurement and analysis of brain activity, enabling finer-scale study of how the brain responds to, processes, and represents visual inputs. Despite these advances, the relationship between the content of visual input from our environment and specific brain responses remains an open question. In particular, mechanistic descriptions of high-level visual processes are difficult to interpret from complex patterns of neural activity. 

In an attempt to understand complex neural activity now available through advanced neuroimaging techniques, high-performing computer vision models have been touted as effective potential models of neural computation \cite{Yamins2016}. This is primarily for two reasons: (1) the origin of these models is linked to the architecture of the primate visual system \cite{LeCun2015} and learning from millions of images; (2) these models achieve high-performance in diverse tasks such as scene recognition, object recognition, segmentation, detection, and action recognition --- tasks defined and grounded in {\em human} judgments of correctness. In this context, a variety of labs have attempted to understand neural data by focusing on the feed-forward hierarchical structure of deep ``convolutional'' neural networks (CNNs) trained on millions of visual images \cite{LeCun2015}. Given a network trained on a dataset for a specific task (e.g., object categorization), one can for a particular brain area: 1) compare the representational properties of different network layers to the representational properties of that neural region; or, 2) use network layers weights to predict neural responses at the voxel (fMRI) or neuron (neurophysiology) level within that neural region. Demonstrating the efficacy of this approach, recent studies have found that higher-level layers of CNNs tend to predict the neural responses of higher-level object and scene regions \cite{Yamins2014, Guclu2015}. Similarly, CNNs have been found to better model human dynamics underlying scene representation \cite{Cichy2016} as compared to more traditional models of scene and object perception (i.e., GIST \cite{Oliva2001} or HMAX \cite{Riesenhuber1999, Serre2005}).

More broadly, there is growing acceptance that computer vision models such as CNNs are useful for understanding biological vision systems and phenomena \cite{Groen2018}. However, computer vision models themselves are still far from approaching human performance and robustness, and there is a growing belief that successful understanding of biological vision will lead to improved computer vision models. As such, integration in both directions will lead to an intertwined approach to biological and computer vision that dates back to \cite{Marr1982}. Indeed, as both models and measurement techniques progress, we come closer to this ideal. However, one of the most significant outstanding challenges for integrating across fields is data \cite{Tarr2016}.

We address this data challenge with the BOLD5000 dataset, a large-scale, slow event-related human fMRI study incorporating 5,000 real-world images as stimuli. BOLD5000 is an order of magnitude larger than any extant slow event-related fMRI dataset, with $\sim20$ hours of MRI scanning per each of four participants. By scaling the size of the image dataset used in fMRI, we hope to facilitate greater integration between the fields of human and computer vision. To that end, BOLD5000 also uniquely uses images drawn from the three most commonly-used computer vision datasets: SUN, COCO, and ImageNet. Beyond standard fMRI analysis techniques, we use both representational similarity analysis \cite{Kriegeskorte2008} and, uniquely, t-distributed stochastic neighbor embedding visualizations \cite{vanDerMaaten2008}, to validate the quality of our data. In sum, we hope that BOLD5000 engenders greater collaboration between the two fields of vision science, fulfilling Marr's dream.

\section*{Methods}

\subsection*{Stimuli}
\subsubsection*{Stimulus Selection}

There are two perspectives to consider for data sharing across biological and computer vision. First, for computer vision, what types of neural data will provide insight or improvement in computer vision systems? Second, for biological vision, what types of images will elicit the best neural data for modeling and understanding neural responses? In this larger context, we suggest that three critical data (i.e., image set) considerations are necessary for success. 

The first data consideration is {\em size}. The general success in modern artificial neural networks can be largely attributed to large-scale datasets \cite{Deng2009}. High-performing models are trained and evaluated on a variety of standard large-scale image datasets. In contrast, although models trained on large-scale datasets have been applied to neural data, the set of images used in these neural studies is significantly smaller --- at best, on the order of a hundred or so distinct images due to time-constrained experimental procedures.

The second data consideration is {\em diversity}. Stimulus sets of constrained size translate into limited diversity for the set of images: the images used in almost every neural study encompass only a small subset of the entire natural image space. For example, despite many studies focusing on object recognition \cite{Khaligh-Razavi2014}, few experiments exceed more than 100 categories across all stimuli. In contrast, image datasets used to train and evaluate artificial neural networks contain thousands of categories, covering a wide range of natural images.

The third data consideration is image {\em overlap}. While multiple recent studies have applied artificial neural networks to the understanding of neural data, the stimulus images used across the two domains have rarely been commensurate with one another. Most significantly, many neural studies use stimuli depicting single objects centered against white backgrounds. In contrast, the images comprising most computer vision datasets contain non-centered objects embedded in realistic, complex, noisy scenes with semantically-meaningful backgrounds. In the instances where studies of biological vision have included more complex images, such as natural scenes, the images were rarely drawn from the same image datasets used in training and testing computer vision models. This lack of overlap across stimuli handicaps the ability to 1) compare the neural data and model representations of visual inputs and 2) utilize neural data in network training or design. 

BOLD5000 tackles these three concerns in an unprecedented slow event-related fMRI study that includes almost 5,000 distinct images. BOLD5000 addresses the issue of size by dramatically increasing the image dataset size relative to nearly all extant human fMRI studies\footnote{Several earlier studies did collect fMRI data while participants viewed real-world movie stimuli \cite{Huth2016,Hasson2004}, in essence, showing thousands of images to participants. However, these images formed events that were necessarily overlapping in time, so no slow event-related analyses were possible. As such, analyses of this kind of ``large-scale'' fMRI data is challenging with respect to disentangling which stimuli gave rise to which neural responses.} --- scaling up by over an order of magnitude. Similarly, BOLD5000 addresses the issues of data diversity and image overlap by including stimulus images drawn from three standard computer vision datasets: scene images from categories largely inspired by Scene UNderstanding (SUN) \cite{Xiao2010}; images directly from Common Objects in Context (COCO) \cite{Lin2014}; and, images directly from ImageNet \cite{Deng2009}. SUN, COCO, and ImageNet, respectively, cover the following image domains: real-world indoor and outdoor scenes; objects interacting in complex, real-world scenes; and objects centered in real-world scenes. These three image datasets cover an extremely broad variety of image types and categories, thereby enabling fine-grained exploration into the neural representation of visual inputs across a wide range of visual features, categories, and semantics.

Furthermore, by including images from computer vision datasets, we address two issues in the computer vision field. Firstly, the datasets we used as well as other computer vision datasets are almost all created by automatically ``scraping'' the web for images, which are then denoised by humans. As a consequence, these ``standard images'' for computer vision have not actually been validated as natural or representative of the visual world around us\footnote{Instead, the generality of CNNs and other modern computer vision models rests on a massive amount of training data, typically covering much of what a model is likely to be tested on in the future.}. In other words, there is little human behavior or neural data on how such images are processed, perceived, and interpreted. BOLD5000 addresses this by providing some metric --- in this case, neural --- of how specific images from these datasets are processed relative to one another. Secondly, the scale of BOLD5000 combined with its overlap with common computer vision datasets enables the possibility of machine learning models {\em training} directly on the neural data associated with images in the training set.

More generally, the scale, diversity, and slow event-related fMRI design of BOLD5000 enables, for the first time, much richer, joint artificial and biological vision models that are critically high-performing in both domains. Beyond the fact that large-scale neural (or behavioral) datasets are necessary for integrating across these two approaches to studying vision, it is also important to note that similarly large-scale neural datasets are equally necessary, in and of themselves, for understanding how complex, real-world visual inputs are processed and represented in the primate brain. In this spirit, BOLD5000 is a publicly-available dataset available at \url{http://BOLD5000.org} (\hyperref[sec:dataCite]{Data Citation 1}).

\begin{figure}[th]
\centering
\includegraphics[width=\textwidth]{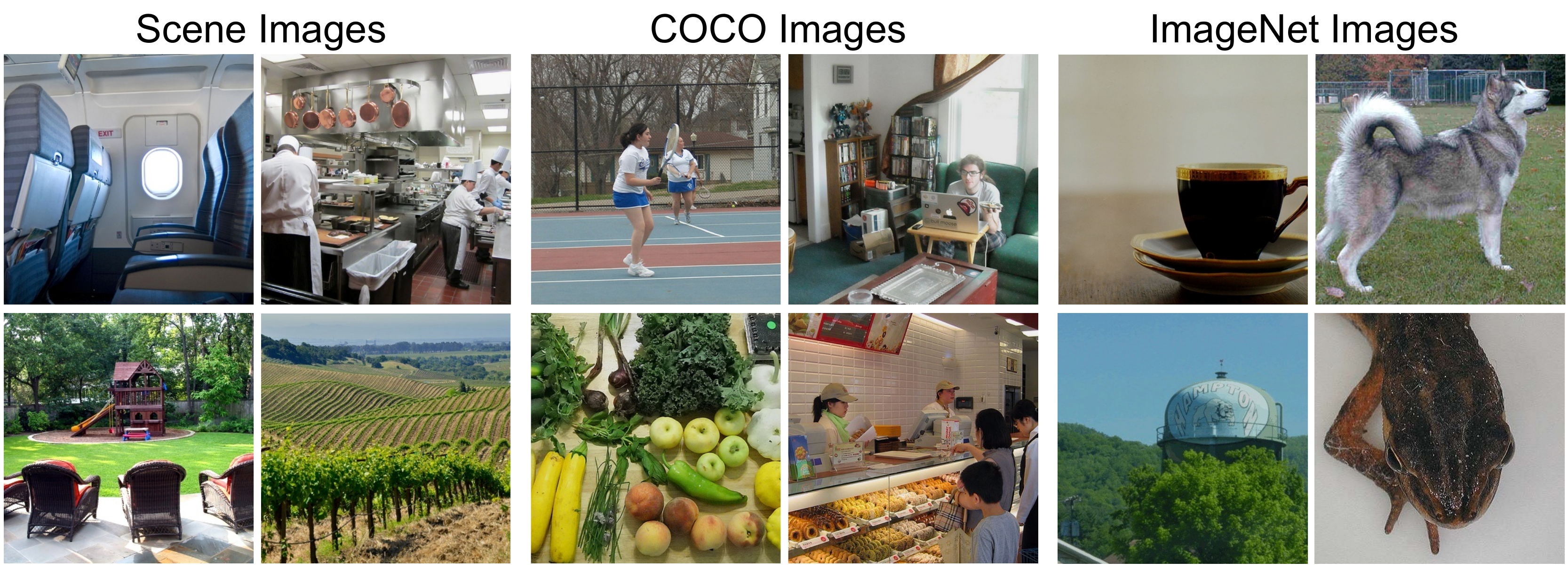}
\caption[Sample Dataset Images]{Sample images from the three computer vision datasets from which experimental image stimuli were selected.}
\label{samples}
\end{figure}

In detail, a total of 5,254 images, of which 4,916 images were unique, were used as the experimental stimuli in BOLD5000. Images were drawn from three particular computer vision datasets because of their prevalence in computer vision and the large image diversity they represented across image categories:

\begin{enumerate}
    \item {\em Scenes.} 1,000 hand-curated indoor and outdoor scene images covering 250 categories, inspired and largely taken from the SUN dataset --- a standard for scene categorization tasks \cite{Xiao2010}. Images in this dataset tended to be more scenic, with less of a focus on any particular object, action, or person. We selected images depicting both outdoor (e.g., mountain scenes) and indoor (e.g., restaurant) scenes.
    
    \item {\em COCO.} 2,000 images of multiple objects from the COCO dataset --- a standard benchmark for object detection tasks \cite{Lin2014}. Due to the complexity of this task, COCO contains images that are similarly complicated and have multiple annotations. Objects tend to be embedded in a realistic context and are frequently shown as interacting with other objects --- both inanimate and animate. COCO is unique because it includes images depicting basic human social interactions.
    
    \item {\em ImageNet.} 1,916 images of mostly singular objects from the ImageNet dataset --- a standard benchmark for object categorization tasks \cite{Russakovsky2015} that is also popular for pre-training CNNs and ``deep'' artificial neural networks \cite{Krizhevsky2012}. Consistent with this task, ImageNet contains images that tend to depict a single object as the focus of the picture. Additionally, the object is often centered and clearly distinguishable from the image background.

\end{enumerate}

Four example images for each of the three datasets are shown in Figure~\ref{samples}.

\subsubsection*{Stimulus Pre-Processing}

To improve the quality of our neural data, we emphasized image quality for our stimulus images by imposing several selection criteria. Basic image quality checks included image resolution, image size, image blurring, and a hard constraint requiring color images only. Additionally, to ensure that sequentially viewing images would not produce neural changes due to image size variation, all images were constrained to be square and of equal size.

To select the 1,000 scene images, we defined 250 unique scene categories (mostly from the SUN dataset) and set a goal of four exemplars per category. We opted not to use the SUN images directly due to a desire to increase the quality of the pictures chosen (i.e., with regard to resolution, watermarks present, etc.). Thus, for each scene category, we queried Google Search with the scene category name and selected images from among the top results according to the above criteria. We only selected images with sufficient size and resolution, then inspected each image to ensure it was clear and free of watermarks. All images were then downsampled to 375 x 375 pixels, the final size for all stimulus images used in this study.

\begin{figure}[th]
% \centering
  \begin{minipage}[b]{0.5\textwidth}
    \includegraphics[width=\textwidth]{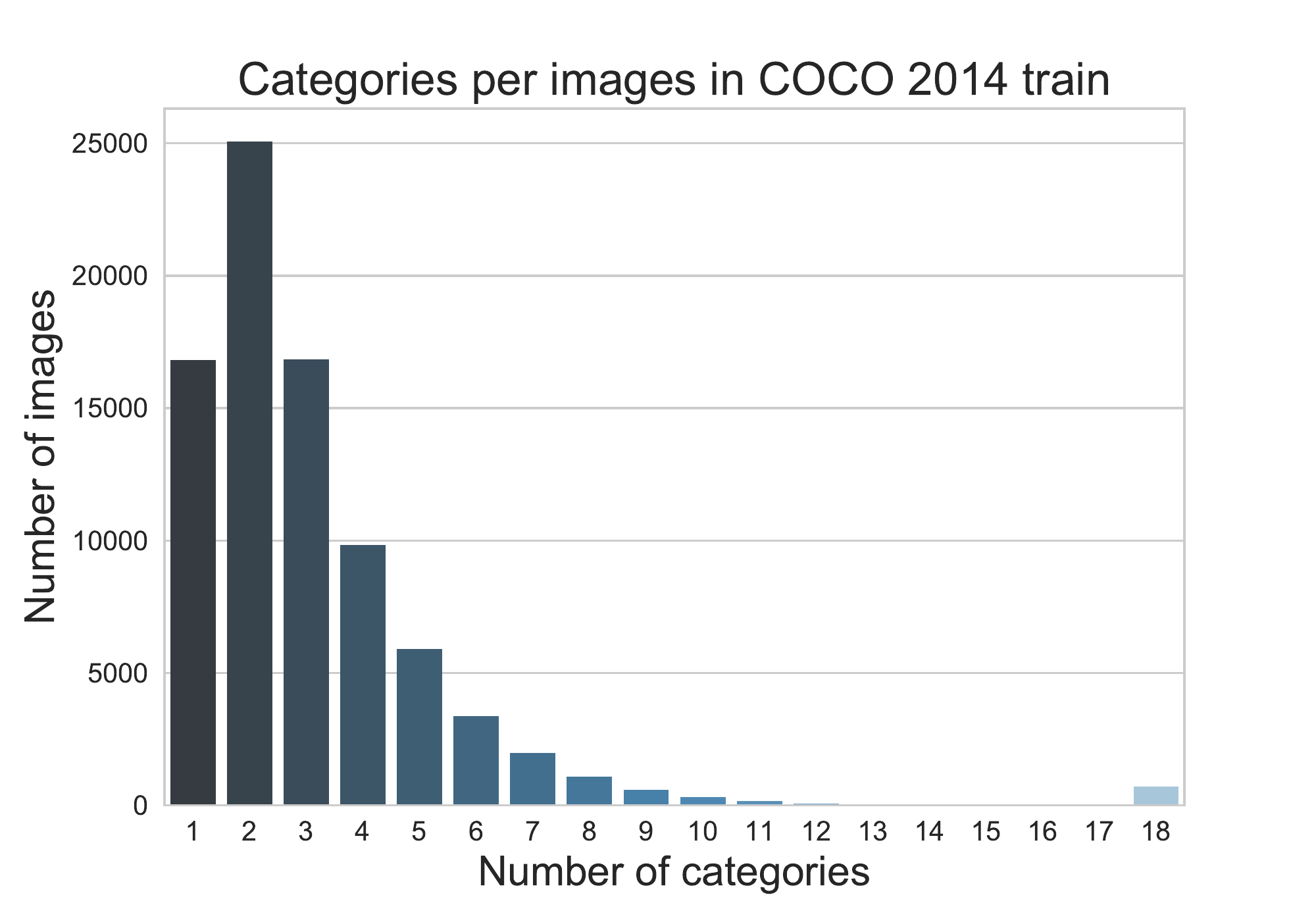}
  \end{minipage}
  \hfill
  \begin{minipage}[b]{0.5\textwidth}
    \includegraphics[width=\textwidth]{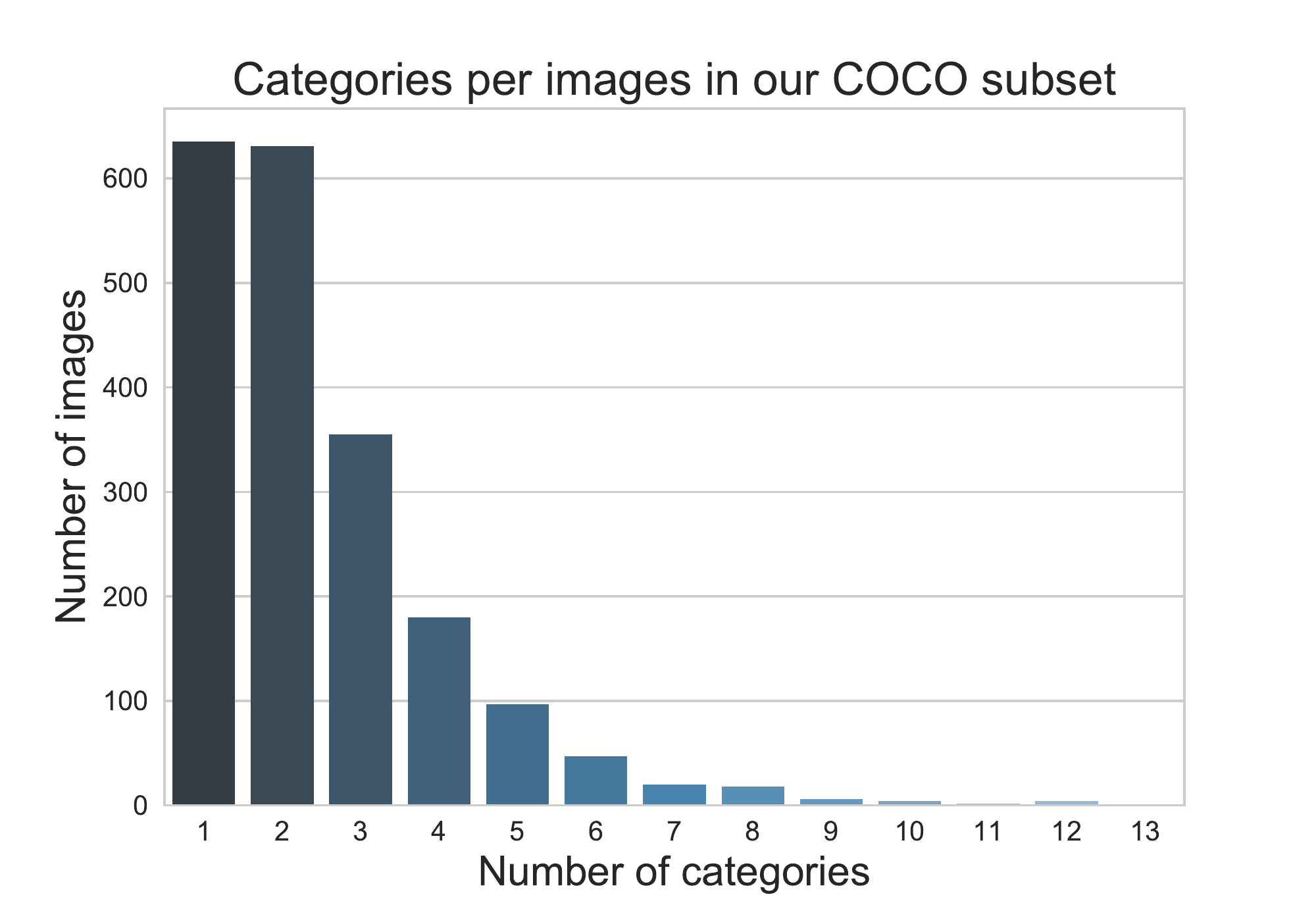}
  \end{minipage}
\caption[COCO statistics for number of categories per image]{Number of images that contain a certain number of categories. Left graph: for the entire COCO 2014 train set, which we sampled from. Right graph: for all 2,000 images selected from the COCO 2014 train set.}
\label{numcats}
\end{figure}

\begin{figure}[th]
  \begin{minipage}[b]{0.5\textwidth}
    \includegraphics[width=\textwidth]{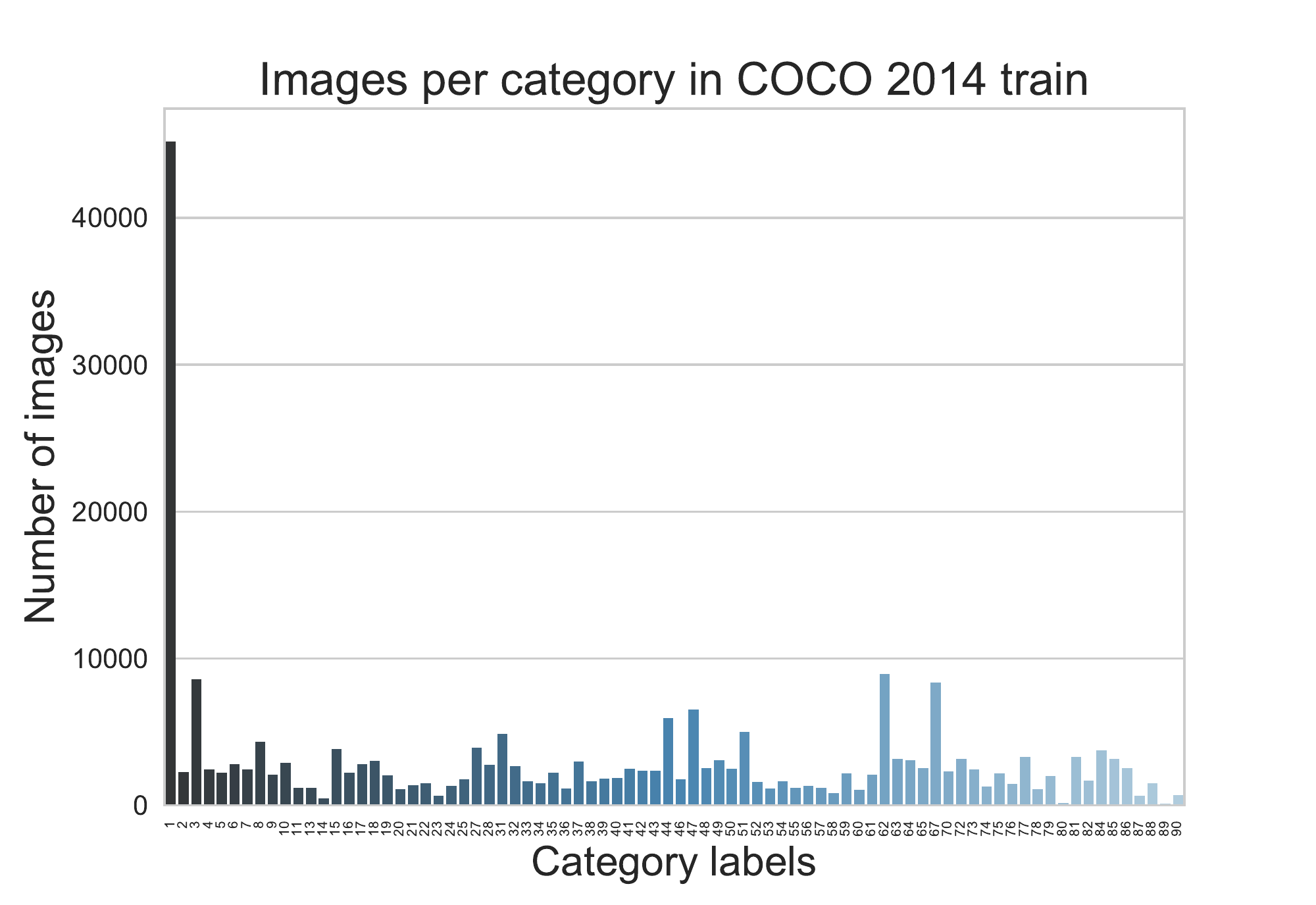}
  \end{minipage}
  \hfill
  \begin{minipage}[b]{0.5\textwidth}
    \includegraphics[width=\textwidth]{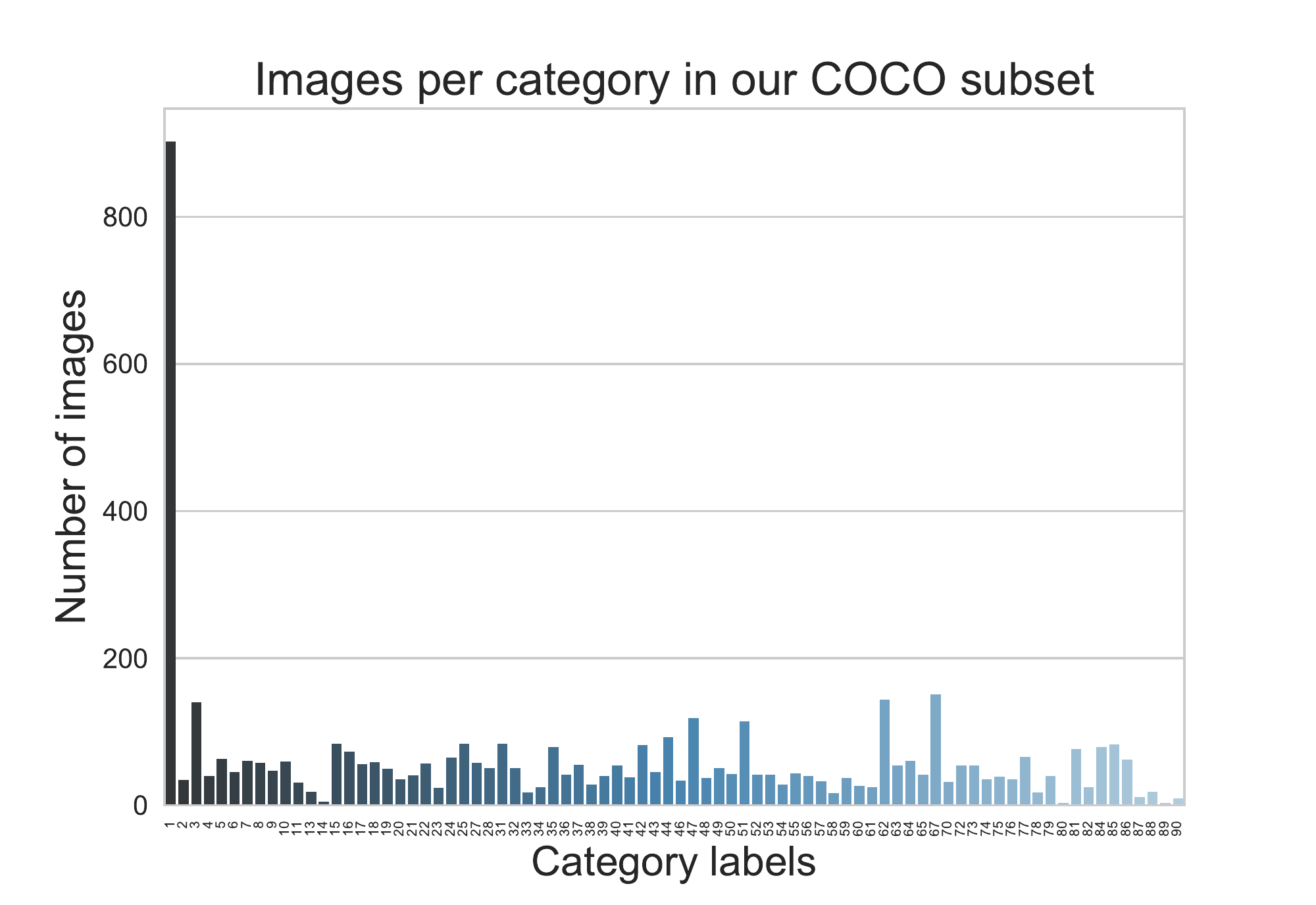}
  \end{minipage}
\caption[COCO statistics for number of images per category]{Number of images that are in each category. Left graph: for the entire COCO 2014 training set, which we sampled from. Right graph: for all 2,000 images selected from the COCO 2014 train set.}
\label{numimgspercat}
\end{figure}

\begin{figure}[th]
  \begin{minipage}[b]{0.5\textwidth}
    \includegraphics[width=\textwidth]{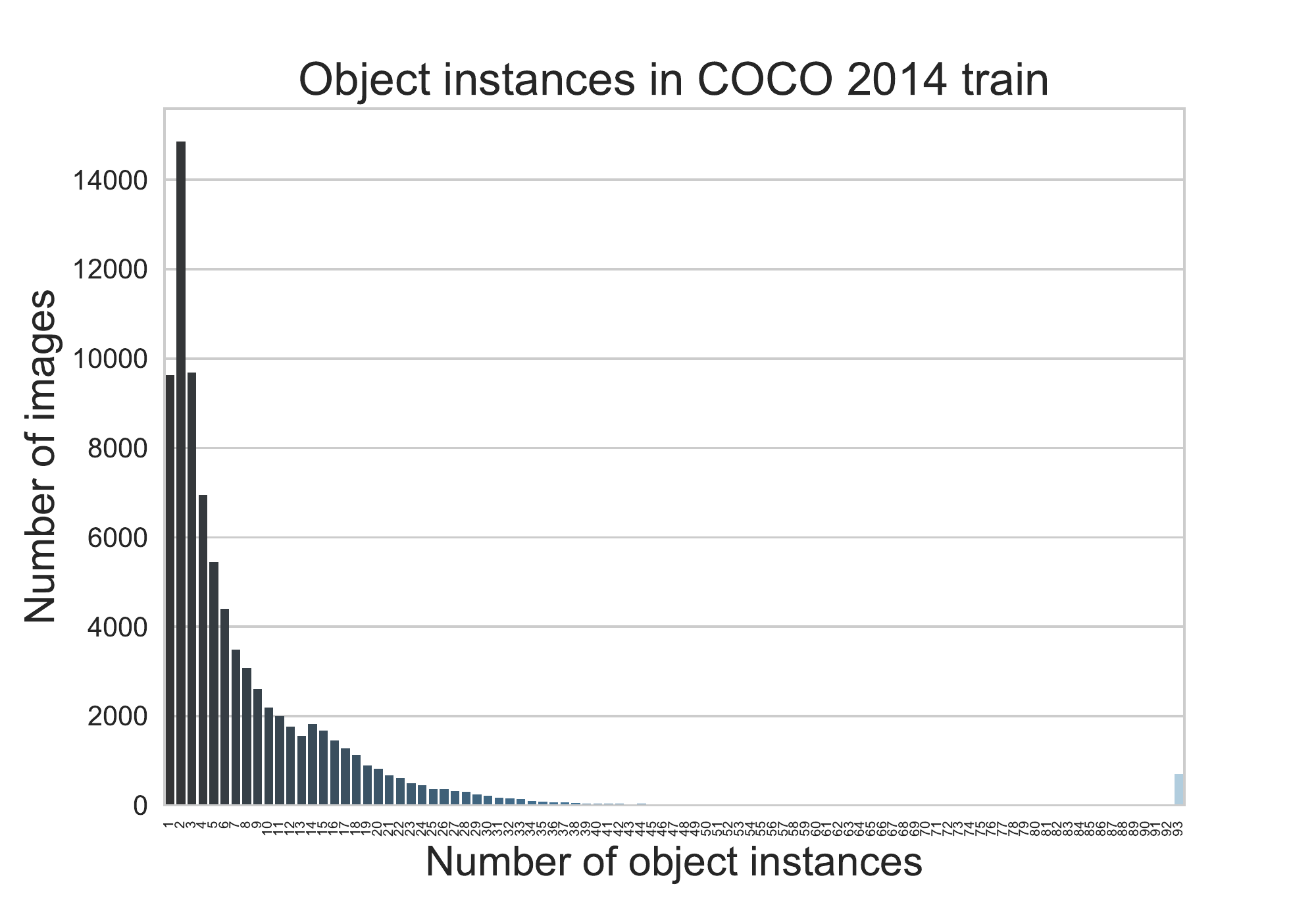}
  \end{minipage}
  \hfill
  \begin{minipage}[b]{0.5\textwidth}
    \includegraphics[width=\textwidth]{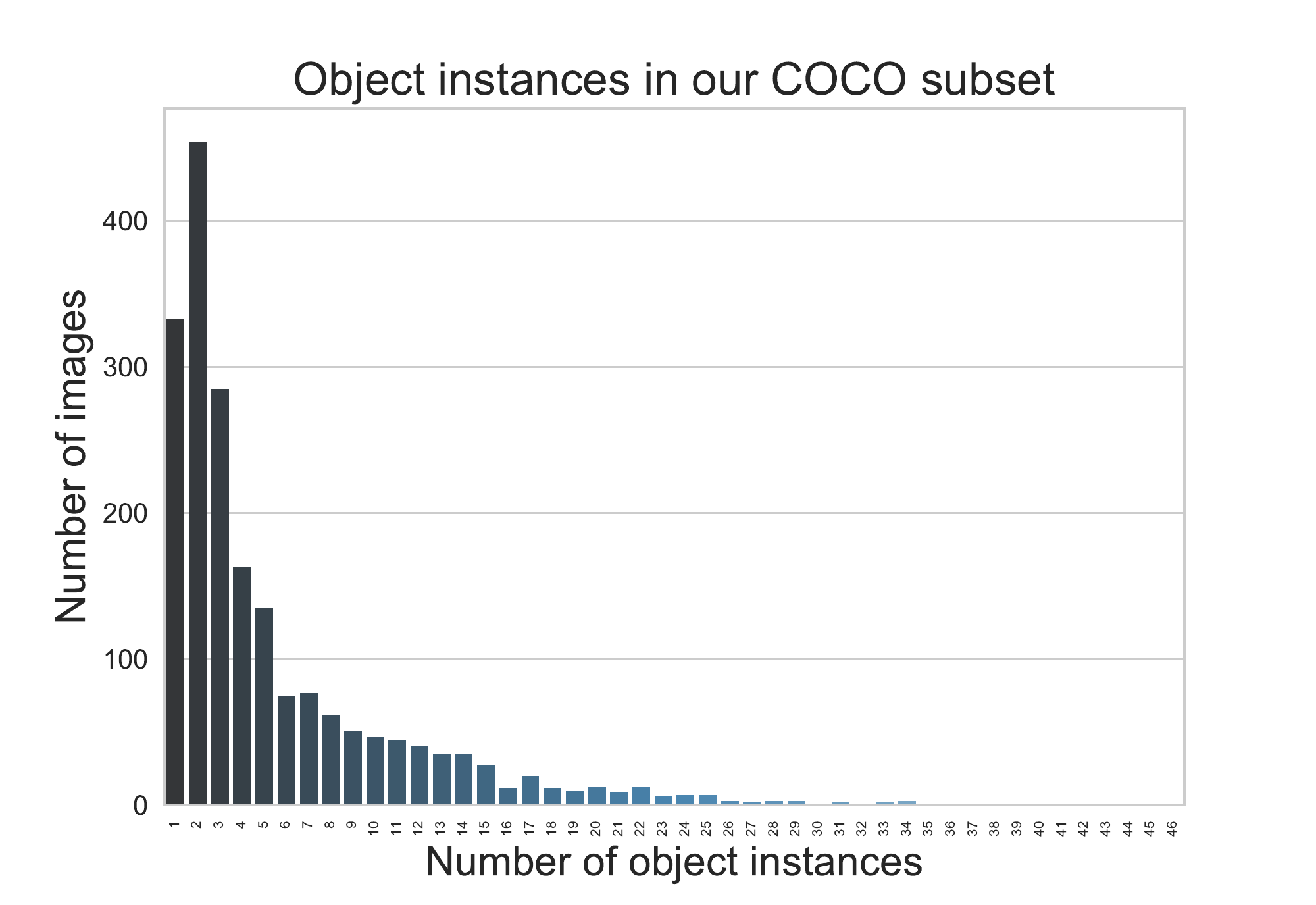}
  \end{minipage}
\caption[COCO statistics for number of object instances in per image]{Number of object instances in each image. Left graph: for the entire COCO 2014 training set, which we sampled from. Right graph: for all 2,000 images selected from the COCO 2014 train set.}
\label{numinsts}
\end{figure}

To select from the COCO images, we sampled 2,000 images from the complete COCO 2014 train dataset. Our goal was to ensure that our chosen images were an accurate representation of the original COCO dataset. Thus, our sampling was structured such that the procedure considered the various annotations that accompany each COCO image. COCO annotations contain 80 object class labels, number of object instances, bounding boxes, and segmentation polygons. Our final 2,000 images adhered to the following criteria: i) the number of categories in our selected images is proportional to that of the training set as shown in Figure~\ref{numcats}; ii) the number of images per category is proportional to that of the training set as shown in Figure~\ref{numimgspercat}; iii) the number of instances per image is proportional to that of the training set as shown in Figure~\ref{numinsts}; iv) the final cropped images contain at least 70\% of the original bounding boxes, where the boxes are counted if there is an intersection over union (area overlap) of at least 50\% between the boxes and the cropped image; v) each image is larger than 375 x 375 pixels. We went through several rounds of sampling, where in each round we randomly sampled according to the above-mentioned criteria before taking a 375 x 375 center crop of the image. Due to the complex, realistic scenes depicted in COCO images, center crops often failed to contain the main image content. Thus, every center cropped image also underwent a manual inspection: if the center crop contained the relevant image content, the crop was retained; if the center crop did not contain the relevant image content, we selected a new region of the image from which to crop. If there was no reasonable crop region, the image was rejected. We repeated this process until 2,000 images had been selected.

To select from the ImageNet images, we used the standard 1,000 class categories in ImageNet for our image selection. However, due to the violent or strong negative nature of some images --- possibly evoking emotional responses --- we removed 42 categories. For each remaining ImageNet category, we randomly selected two exemplars per category from the training set that satisfied our image size and resolution criteria. With 958 categories and two exemplars per category, we obtained a total of 1,916 ImageNet images. In light of ImageNet images having varying sizes and resolutions, we only considered images that were larger than 375 x 375 pixels before taking a 375 x 375 center crop. For all randomly sampled center crops, we manually inspected each image to ensure that the crop did not exclude a large portion of the image content and that the image resolution was sufficiently high. We repeated this process until two exemplars per category had been selected.

Finally, we considered the RGB and luminance distribution across all of the selected images. Because visual brain responses are influenced by image luminance, we attempted to render our images invariant to this factor. To this end, for each image we calculated its hue, saturation, and value (HSV). This value represents the brightness of the image. The average brightness per image was determined, and we calculated the difference between average brightness and gray brightness. All values were then multiplied in the image by this new scale --- a process known as gray world normalization. This process ensured that luminance was as uniform as possible across all images.

In order to examine the effect of image repetition, we randomly selected 112 of the 4,916 distinct images to be shown four times and one image to be shown three times to each participant. These 113 images were selected such that the image dataset breakdown was proportionally to that of the 4,916 distinct images. Specifically, $1/5$ of the images were scene images, $2/5$ of the images were COCO images, $2/5$ of the images were ImageNet images. When these image repetitions are considered, we have a total of 5,254 image presentations shown to each participant (4,803 distinct images + $4 \times 112$ repeated images + $3 \times 1$ repeated image). For CSI3 and CSI4, a small number of repetitions varied from 2-5 times. 

%--------------------------------------------------------------------
\subsection*{Experimental Design}

\subsubsection*{General Procedure}
fMRI data were collected from a total of four participants (referred to as CSI1, CSI2, CSI3, CSI4), with a full dataset collected for three of the four participants (see Participant Selection). A full dataset was collected over 16 MRI scanning sessions, where 15 were functional sessions acquiring task relevant data, and the remaining session comprised of collecting high resolution anatomical and diffusion data. For CSI4, there were 9 functional sessions, with an additional anatomical session. Participants were scanned using custom headcases (CaseForge Inc.) to reduce head movement and maintain consistent head placement and alignment across sessions. 

For all participants, each of the functional scanning sessions was 1.5 hours long: 8 sessions had 9 image runs and 7 sessions had 10 image runs. In the sessions with only 9 image runs, we included an additional functional localizer run at the end of the session, providing a total of 8 localizer runs across 15 sessions. The functional localizer runs were used to independently define regions of interest for subsequent analyses. Over the course of the 15 functional sessions, all 5,254 image trials (3,108 for CSI4) were presented. During each functional session the participant sequentially viewed 5 functional scans with less than a 1 minute break in between each scan. Anatomical scans were then collected over the course of approximately 4 minutes, during which the participants were allowed to close their eyes or view a movie. Finally, the last sequential 5 functional scans (or 4 if a localizer was run at the end of the session) were run. After all scans, the participant filled out a questionnaire (Daily Intake) about their daily routine, including: current status regarding food and beverage intake, sleep, exercise, ibuprofen, and comfort in the scanner.

\subsubsection*{Experimental Paradigm for Scene Images}
The following run and session details apply for each participant. Each run contained 37 stimuli. In order for the images in a given run to accurately reflect the entire image dataset, the stimuli in each run were proportionally the same as the overall dataset: roughly $ $1/5$^{th}$ scene images, $ $2/5$^{th}$ COCO images, and $ $2/5$^{th}$ ImageNet images. Of the 37 images in each run, roughly 2 were from the set of repeated images. For the 35 unique images per run, 7 were scene images, 14 were COCO images, and 14 were ImageNet images. However, because the total number of images does not divide evenly into 7's, some sessions contained a slightly unbalanced number of image categories by a factor of 1 image. Finally, the presentation order of the stimulus images was randomly and uniquely determined for each participant. These presentations orders were fixed before the start of any experimental sessions for all four participants. All stimuli (both scenes and localizer images) were presented using Psychophysics Toolbox Version 3 (``Psychtoolbox'')\cite{Brainard1997}
running under Matlab (Mathworks, Natick, MA) on a Systems76 computer running Ubuntu presented via 24-inch MR compatible LCD display (BOLDScreen, Cambridge Research Systems LTD., UK) mounted at the head end of the scanner bore. Participants viewed the display through a mirror mounted on the head coil. All stimulus images were 375 x 375 pixels and subtended approximately 4.6 degrees of visual angle.  

The following image presentation details apply for each run, each session, and each participant. A slow event-related design was implemented for stimulus presentation in order to isolate the blood oxygen level dependent (BOLD) signal for each individual image trial. At the beginning and end of each run, centered on a blank, black screen, a fixation cross was shown for 6~sec and 12~sec, respectively. Following the initial fixation cross, all 37 stimuli were shown sequentially. Each image was presented for 1~sec followed by a 9~sec fixation cross. Given that each run contains 37 stimuli, there was a total of 370~sec of stimulus presentation plus fixation. Including the pre- and post-stimulus fixations, there were a total of 388~sec (6~min 28~sec) of data acquired in each run. 

For each stimulus image shown, each participant performed a valence judgment task, responding with how much they liked the image using the metric: ``like'', ``neutral'', ``dislike''. Responses were collected during the 9~sec interval comprising the interstimulus fixation, that is, subsequent to the stimulus presentation, and were made by pressing buttons attached to an MRI-compatible response glove on their dominant right hand.

\subsubsection*{Experimental Paradigm for the Functional Localizer}
As mentioned above, 8 functional localizers (6 for CSI4) were acquired throughout the 15 functional sessions. These localizers were run at the end of the session for that day. The functional localizer included three conditions: scenes, objects, and scrambled images. The scene condition included images depicting indoor and outdoor environments and were non-overlapping with the 4,916 scenes used in the main experiment. The object condition included objects with no strong contextual association (i.e. weak contextual objects \cite{Bar2003}). Scrambled images were generated by scrambling the Fourier transform of each scene image. There were 60 color images in each condition. Images were presented at a visual angle of 5.5 degrees. 

Each run was designed in a block design format. Each block had 16 trials, with a stimulus duration of 800 ms, and a 200 ms ISI. Of the 16 trials, there were 14 unique images, with 2 images repeating. Participant's task was to press a button if an image immediately repeated (i.e., a one-back task). Between task blocks there were six seconds of fixation. Each run started and ended with 12 seconds of fixation. There were 12 blocks per run, with 4 blocks per condition.   

\subsubsection*{MRI Acquisition Parameters}
MRI data were acquired on a 3T Siemens Verio MR scanner at the Carnegie Mellon University campus using a 32-channel phased array head coil. 

\textit{Functional Images and Fieldmaps:} Functional images were collected using a T2*-weighted gradient recalled echo echoplanar imaging multi-band pulse sequence (cmrr\_mbep2d\_bold) from the University of Minnesota Center for Magnetic Resonance Research (CMRR) \cite{Moeller2010, Feinberg2010}. Parameters: 69~slices co-planar with the AC/PC; in-plane resolution = 2~x~2~mm; 106~x~106 matrix size; 2~mm slice thickness, no gap; interleaved acquisition; field of view = 212~mm; phase partial Fourier scheme of 6/8; TR = 2000~ms; TE = 30~ms; flip angle = 79~degrees; bandwidth = 1814~Hz/Px; echo spacing = .72~ms; excite pulse duration = 8200~microseconds; multi-band factor = 3; phase encoding direction = PA; fat saturation on; advanced shim mode on. Data were saved both with and without pre-scan normalization filter applied. During functional scans, physiological data was also collected using wireless sensors: heart rate was acquired using a Siemens photoplethysmograph attached to the participant's left index finger; respiration was acquired using a respiratory cushion held to the participant's chest with a belt and connected via pressure hose to a Siemens respiratory sensor. In the middle of each scanning session, three sets of scans were acquired for use for geometric distortion correction. The three options were provided to allow researchers to use the scans that work best in their processing pipeline. Options: 1) A three volume run with phase encoding (PA) exactly matching the functional scan (opposite phase encoding achieved using the "Invert RO/PE polarity" option). 2) Two pairs of three volume spin-echo runs with opposite phase encoding, one with partial Fourier 6/8, one without, both using the cmrr\_mb3p2d\_se pulse sequence. Non partial Fourier spin-echo parameters: geometry and resolution matched to functional parameters; TR = 9240~ms; TE = 81.6~ms; multi-band factor = 1. 3) Partial Fourier spin-echo parameters: geometry and resolution matched to functional parameters; TR = 6708~ms; TE = 45~ms; phase partial Fourier scheme of 6/8; multi-band factor = 1. Opposite phase encoding achieved using "Invert RO/PE polarity" option.   

\textit{Anatomical and Diffusion Images:} A T1 weighted MPRAGE scan, and a T2 weighted SPACE scan using Siemens pulse sequences were collected for each participant. MPRAGE parameters: 176 sagittal slices; 1~mm isovoxel resolution; field of view = 256~mm; TR = 2300~ms; TE = 1.97~ms; TI = 900~ms: flip angle = 9 degrees; GRAPPA acceleration factor = 2; bandwidth = 240~Hz/Px. SPACE parameters: 176 sagittal slices; 1~mm isovoxel resolution; field of view = 256~mm; TR = 3000~ms; TE = 422~ms; GRAPPA acceleration factor = 2; bandwidth = 751~Hz/Px; echo spacing = 3.42~ms. Participants' faces were removed from the MPRAGE and SPACE scans to protect privacy using Pydeface \cite{pydeface}. Two diffusion spectrum imaging (DSI) scans were acquired for each participant using the cmrr\_mbep2d\_diff sequence from CMRR. Parameters: geometry and resolution matched to functional parameters; diffusion spectrum imaging sampling scheme; 230 directions; phase partial Fourier scheme of 6/8; TR = 3981~ms; TE = 121~ms; maximum b-value = 3980~s/mm$^2$; bipolar acquisition scheme; AP phase encoding direction. A second scan matching all parameters but with PA phase encoding (achieved using the “Invert RO/PE Polarity” option) was also acquired.

\subsection*{Data Analyses}

\subsubsection*{fMRI Data Analysis}

All fMRI data were converted from DICOM format into Brain Imaging Data Structure (BIDS http://bids.neuroimaging.io/) using a modified version of dcm2bids \cite{dcmBIDS}. Data with the pre-scan normalization filter applied were used for all analyses.  Data quality was assessed and image quality metrics were extracted using the default pipeline of MRIQC \cite{Esteban2017}.  

Results included in this manuscript come from preprocessing performed using FMRIPREP 1.1.4 \cite{Esteban2017, zenodo}, a Nipype \cite{Gorgolewski2011, gorgo} based tool. Each T1w (T1-weighted) volume was corrected for INU (intensity non-uniformity) using N4BiasFieldCorrection v2.1.0 \cite{Tustison2010} and skull-stripped using antsBrainExtraction.sh v2.1.0 (using the OASIS template). Brain surfaces were reconstructed using recon-all from FreeSurfer v6.0.1 \cite{Dale1999}, and the brain mask estimated previously was refined with a custom variation of the method to reconcile ANTs-derived and FreeSurfer-derived segmentations of the cortical gray-matter of Mindboggle \cite{Klein2017}. Spatial normalization to the ICBM 152 Nonlinear Asymmetrical template version 2009c \cite{Fonov2009} was performed through nonlinear registration with the antsRegistration tool of ANTs v2.1.0 \cite{AVANTS2008}, using brain-extracted versions of both T1w volume and template. Brain tissue segmentation of cerebrospinal fluid (CSF), white-matter (WM) and gray-matter (GM) was performed on the brain-extracted T1w using fast \cite{Zhang2001} (FSL v5.0.9).

Functional data were motion corrected using mcflirt (FSL v5.0.9 \cite{Jenkinson2002}). Distortion correction was performed using an implementation of the Phase Encoding POLARity (PEPOLAR) technique using 3dQwarp (AFNI v16.2.07 \cite{Cox1996}). This was followed by co-registration to the corresponding T1w using boundary-based registration \cite{Greve2009} with 9 degrees of freedom, using bbregister (FreeSurfer v6.0.1). Motion correcting transformations, field distortion correcting warp, and BOLD-to-T1w transformation were concatenated and applied in a single step using antsApplyTransforms (ANTs v2.1.0) using Lanczos interpolation.

Many internal operations of FMRIPREP use Nilearn \cite{Abraham2014}, principally within the BOLD-processing workflow. For more details of the pipeline see \url{https://fmriprep.readthedocs.io/en/latest/workflows.html}. All reports of the fMRIPrep analysis are publicly available, see Data Usage. 

After preprocessing, the data for the 5,254 images were analyzed and extracted using the following steps. Data from each session, (i.e., including 9 or 10 runs, depending on session) were entered into general linear model (GLM), where nuisance variables were regressed out of the data. There were a total of nine nuisance variables including 6 motion parameters estimates resulting from motion correction, the average signal inside the cerebral spinal fluid mask and separately for the white matter mask across time, as well as global signal within the whole-brain mask. All of the nuisance variables were confounds extracted in the fMRIPREP analysis stream. A regressor for each run of the session was used in the GLM. In addition, a high pass filter of 128s was applied to the data. The residual time series of each voxel within a region of interest (see details below) were extracted, demeaned across all image presentations, and used for all subsequent analyses.

\subsubsection*{Functional Localizer Analysis}
Participants CSI1, CSI2, and CSI3 all had eight functional localizer runs scattered throughout the 15 functional sessions, with a maximum of one localizer run per session. CSI4 had six functional localizer runs scattered throughout the 9 functional sessions. All localizer data post fMRIPREP preprocessing were analyzed within the same general linear model using a canonical hemodynamic response function implemented using SPM12 \cite{spm}. All 9 nuisance regressors mentioned above, plus regressors for each run, and a high pass filter were implemented in the model. Three conditions were modeled: scenes, objects, and scrambled images.

\subsubsection*{Region of Interest (ROI) Analyses}
All ROI analyses were performed at the individual level using the MarsBaR toolbox \cite{Brett2002} and analyzed within native space on the volume. Scene selective regions of interest (the parahippocampal place area, PPA; the retrosplenial complex, RSC, and the occipital place area, OPA) were all defined using the contrast of scenes compared with objects and scrambled. Although the functional localizer was not specifically designed to locate object selective areas, a lateral occipital complex (LOC) was successfully defined by comparing objects to scrambled images. Finally, an early visual (EarlyVis) ROI was defined by comparing the scrambled images to baseline, and the cluster most confined to the calcarine sulcus was used. However, since retinotopy was not used to define the EarlyVis ROI, and an anatomical boundary was not applied, this region sometimes extends beyond what is classically considered V1 or V2. A threshold of using a family-wise error correction of p < .0001 (or smaller), and k = 30 was used for all ROIs. In cases where including all localizer runs resulted in clusters of activity too big for the intended ROI (e.g., the PPA ROI connected to the RSC ROI), a reduced number of runs were included in the contrast. For the scene selective and object selective regions, this included using just the first, middle, and last run. Across all participants, the EarlyVis ROI was only defined using the first run due to massive amount of activity produced from this contrast. Using these ROIs defined from the localizer, the data of the 5,254 scenes were then extracted across the timecourse of the trial presentation (5TRs; 10sec). We analyzed the data for each timepoint (TR1 = 0-2s, TR2 = 2-4s, TR3 = 4-6s, TR4 = 6-8s, TR5 = 8-10s). 

To demonstrate the timecourse of the data, the time series was extracted for each ROI and averaged across all image presentations (e.g., N = 5,254), see Technical Validation. For CSI1, CSI2, and CSI3 the timecourse shows that activity peaked across TR3 and TR4 (4-8s). Due to this pattern of results, all subsequent data analysis uses an average of the data from TR3 and TR4. For CSI4, the timecourse demonstrates a peak of activity most confined to TR3. Thus, only data from TR3 for CSI4 was used for subsequent analyses. 

Finally, data from extracted ROIs were demeaned for normalization purposes. Data were demeaned by each voxel by taking the mean of that voxel across all image presentations and subtracting it from each sample data point.

\subsection*{Participant Selection}

Participants were recruited from the pool of students enrolled in graduate school at Carnegie Mellon University. Due to the multi-session nature of the study, we specifically recruited participants who were familiar with MRI procedures and believed themselves capable of completing all scanning sessions with a minimum effect on data quality (e.g., low movement, remaining awake, etc.). This requirement necessarily limited the number of individuals willing and capable of participating in this study. In this context, our study included four participants: three participants (CSI1, CSI2, CSI3) with a full set of data (all 16 sessions), and a fourth participant (CSI4) who only completed 10 sessions due to discomfort in the MRI. Participant demographics are as follows: CSI1 - male, age 27, right handed; CSI2 - female, age 26, right handed; CSI3 - female, age 24, right handed; CSI4 - female, age 25, right handed. Participants all reported no history of psychiatric or neurological disorders and no current use of any psychoactive medications. Participants all provided written informed consent and were financially compensated for their participation. All procedures followed the principles in the Declaration of Helsinki and were approved by the Institutional Review Board of Carnegie Mellon University.

\begin{table}[h!]
\centering 
\includegraphics[height=.73\textheight]{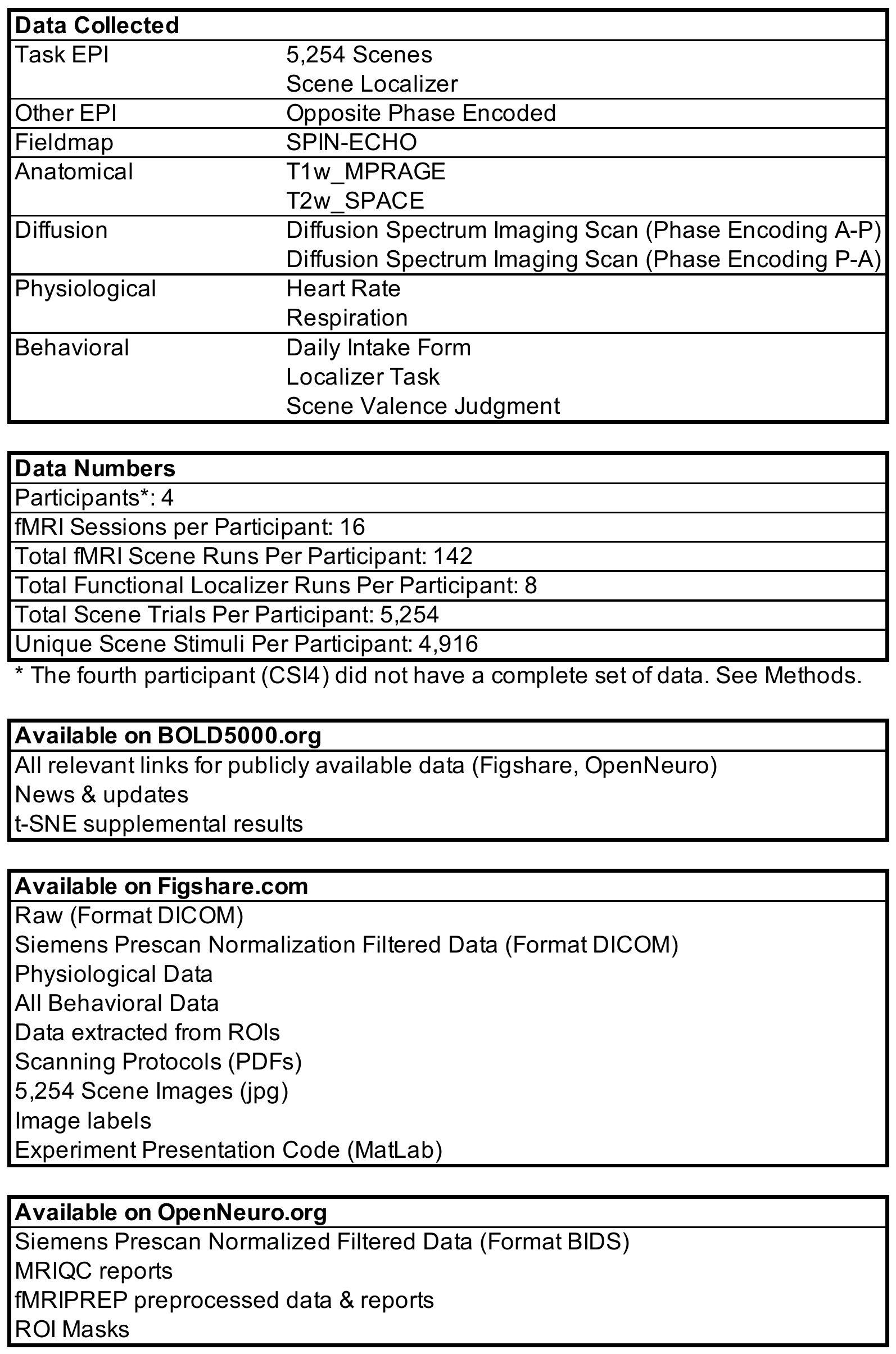}
\caption{A list of data provided.}
\label{data}
\end{table}

\subsection*{Code Availability}

The complete set of Psychtoolbox Matlab scripts for running this study are available for download at \url{http://scripts.bold5000.org}. The Psychtoolbox Version 3 (PTB-3) Matlab toolbox and documentation are both available for download at \url{http://psychtoolbox.org}. The complete set of images used as stimuli are available for download at \url{http://images.bold5000.org}, packaged in the file \url{BOLD5000_Stimuli.zip}.

\section*{Data Records}
\label{sec:datarec}
We have publicly released BOLD5000 online. As seen in Table~\ref{data}, we have provided a comprehensive list of collected data and the various stages of analyzed data we have made available. All relevant links and information can be found at \url{http://BOLD5000.org}.

\section*{Technical Validation}

\begin{table}[h!]
\centering
\includegraphics[width=\textwidth]{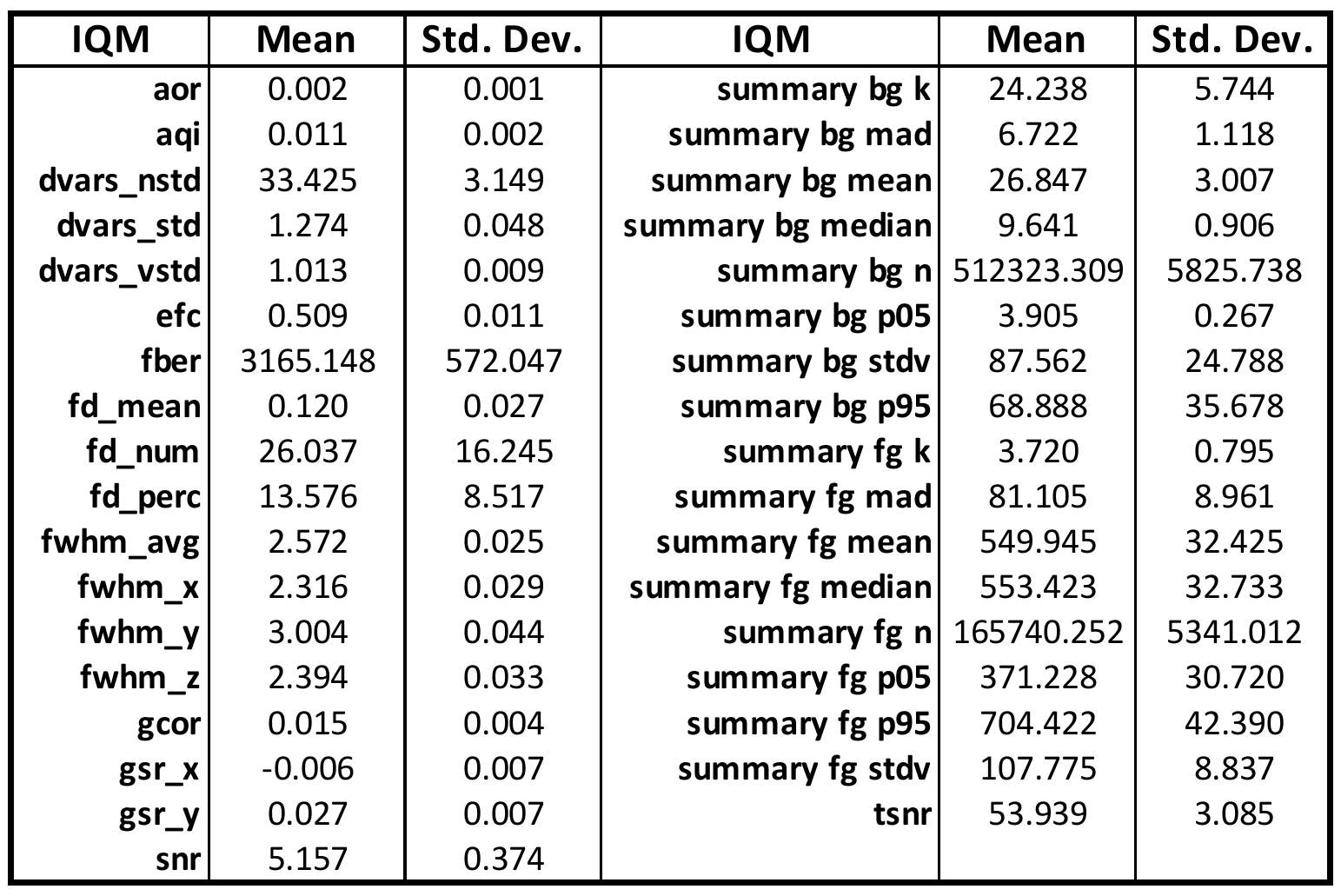}
\caption{A list of the image quality metrics (IQMs) produced as a result of the MRIQC analysis. Each mean and standard deviation is calculated across all runs of all participants (CSI1,2,3 N = 150; CSI4 N = 90). For information about each IQM please refer to \url{http://mriqc.org}.}
\label{IQM}
\end{table}

\subsection*{Data Quality} 

\begin{figure}[h!]
\centering
\includegraphics[width=0.89\textwidth]{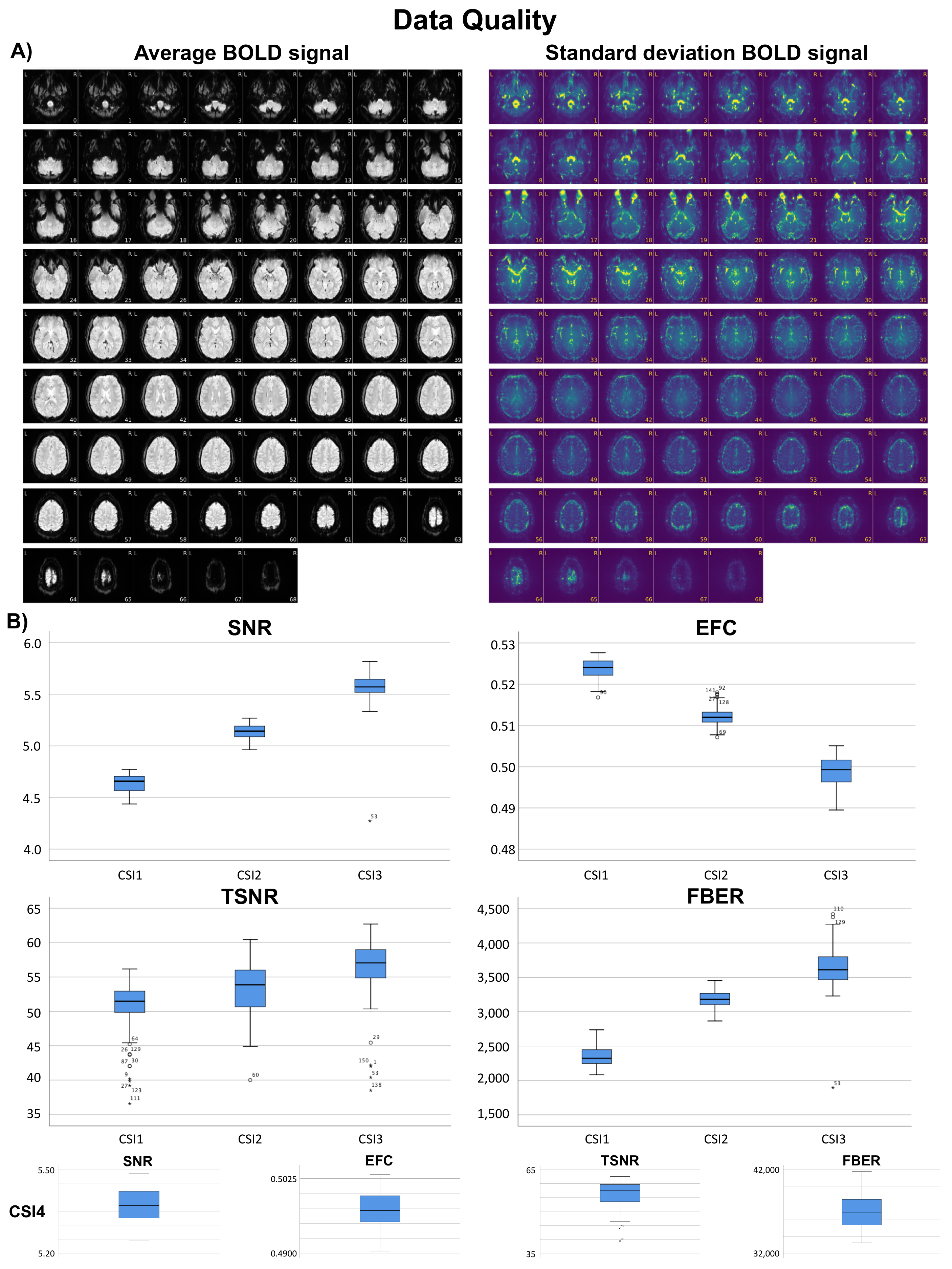}
\caption[Sample Dataset Images]{ A) A single participant's (CSI1) average BOLD signal (Left) and standard deviation of BOLD signal (Right) for a single run (Session 1, Run 1). B) Boxplots of representative IQMs for each run of each participant (N = 150 per participant). SNR - signal to noise ratio; higher values are better. TSNR - temporal signal to noise ratio; higher values are better. EFC - entropy focus criterion - a metric indicating ghosting and blurring induced by head motion; lower values are better. FBER - foreground to background energy ration; higher values are better. Data for CSI4 is separate due to the different number of datapoints (N = 90).}
\label{DQ}
\end{figure}

To provide measures describing the quality of this dataset, the data were analyzed using MRIQC \cite{Esteban2017}. MRIQC is an open-source analysis program that provides image quality metrics (IQMs) in an effort to provide interoperability, uniform standards, and to assess reliability of a dataset, ultimately with the goal to increase reproducibility. MRIQC analyzes each run (in this case, 150 per participant) and provides IQMs for each run, as well as a figure of the average BOLD signal in the run as well as the standard deviation for each run, see Figure~\ref{DQ} for an example from CSI1 – session 1, run 1. For a full report of each run for each participant, please find the complete analysis available on OpenNeuro.org (see \hyperref[sec:datarec]{Data Records}). Figure~\ref{DQ} also demonstrates the stability of the data across all 150 runs by showing the variance across four representative IQMs across all runs. Table ~\ref{IQM} provides the averages and standard deviations across all participants for all measures from the MRIQC analysis. These measures are on par, if not better, than other studies that have provided IQMs (e.g., see \url{http://mriqc.org}). 

\subsection*{Design Validation}

\begin{figure}[th]
\centering
\includegraphics[scale=0.64]{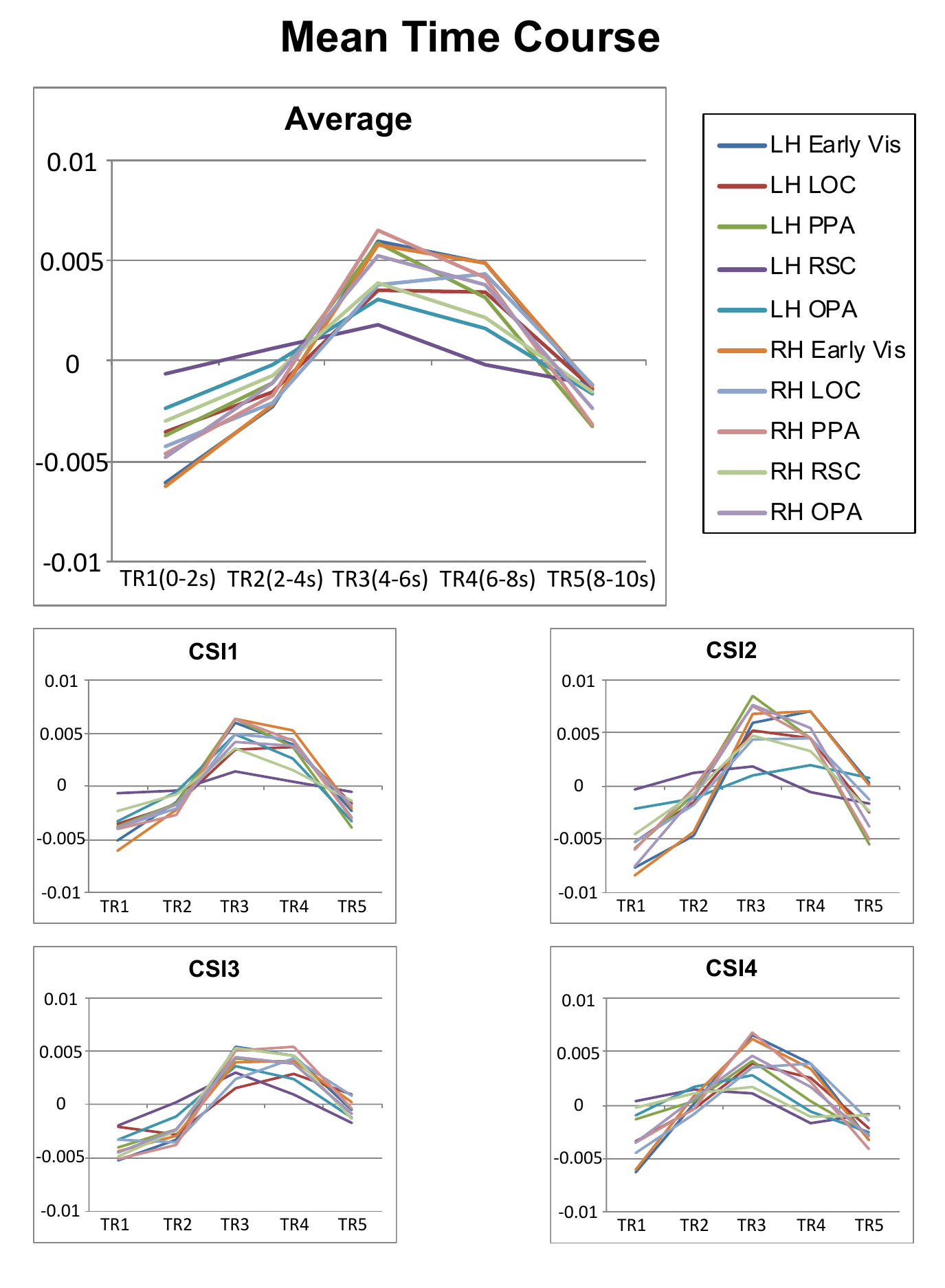}
\caption[timecourse]{Mean time course across all stimulus presentations for each region of interest.}
\label{timecourse}
\end{figure}

In this study, our goal was to make the data accessible without requiring advanced fMRI analysis tools to understand the data. Therefore, we ensured that each trial was isolated from neighboring trials. In this respect, we chose a slow event-related design in which the stimulus was presented for one second, and a fixation cross was presented for nine seconds between stimulus presentations. As a result of using a slow event-related design, the extracted timecourse shows the hemodynamic response peaked around 6 seconds post stimulus onset, and returned to baseline before the next stimulus was presented. Using this design, there was no bleed over from neighboring trials and no need for deconvolving the signal. Figure~\ref{timecourse} shows the timecourse averaged across all stimulus trials for each region of interest. As can be seen in the figure, the timing of the design allowed the BOLD signal to peak and return to baseline.

\subsection*{Data Validation}
\subsubsection*{Repeated Stimulus Images}
\begin{figure}[th]
% \centering
\includegraphics[width=\textwidth]{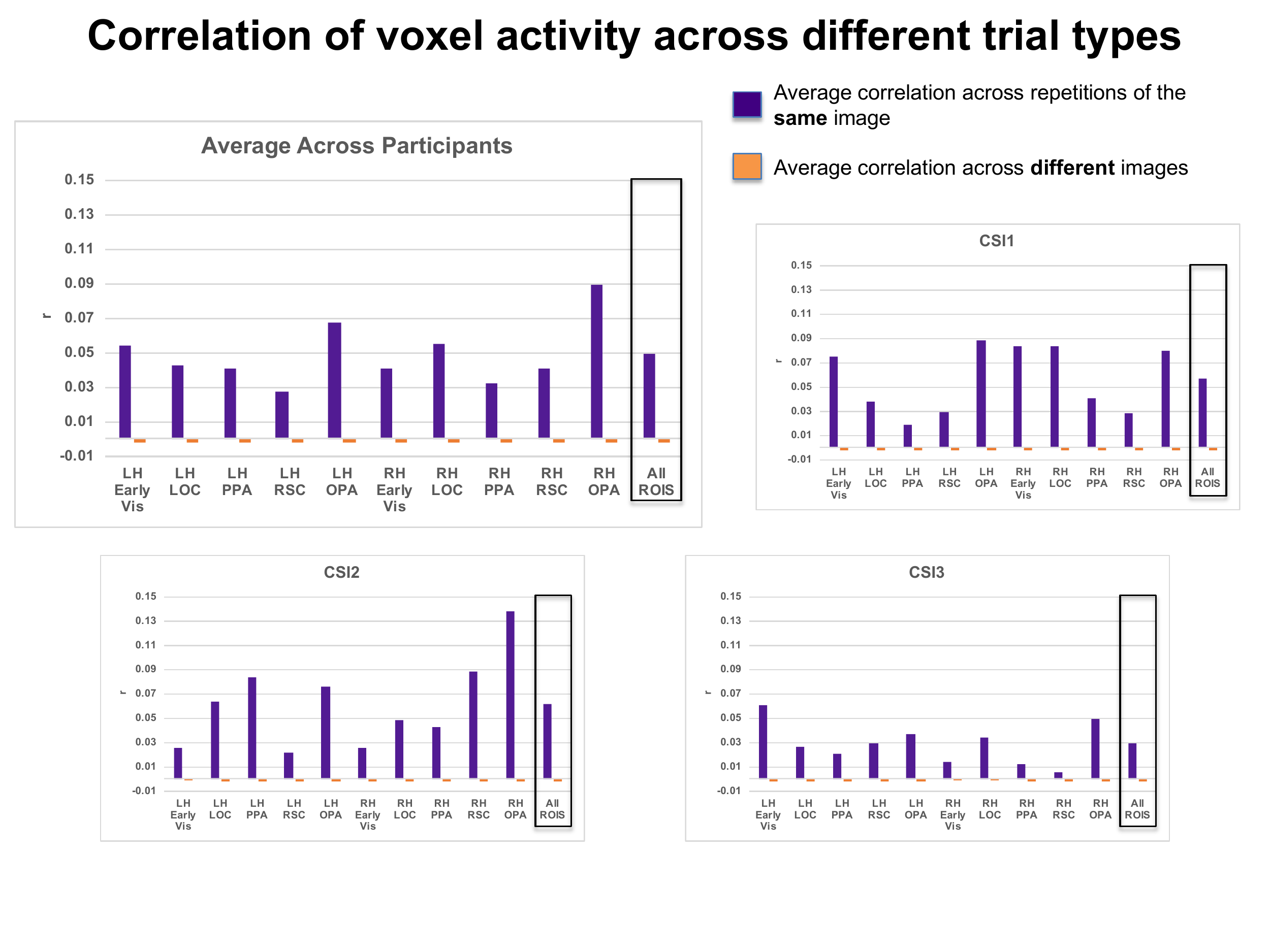}
\caption[Correlations]{Correlations of BOLD signal across voxels within each region of interest for repetitions of the same image in comparison with the correlation across different images. A box surrounds the columns representing the averages across all ROIs.}
\label{corr}
\end{figure}

As noted above, a majority of our images, 4,803 were only presented to the participant on a single trial during the entire 15 functional sessions. However, 113 images had repeated presentations (3+) over the 15 functional sessions. We implemented this design to assess the signal to noise ratio in the data.  Note that repeated stimulus means that we have 4 unique neural representations for the same stimulus. Since the stimulus is the same, the neural representations are expected to be the same, with the exclusion of noise and session to session variance. Thus, we leverage the extra neural representations to our advantage. To demonstrate this, the reliability of the BOLD pattern across voxels within a given ROI for each repetition of a given image was examined. In this analysis, we hypothesize that the correlation of the pattern of BOLD activity across the repetitions of the same image should be considerably higher than the correlation of the patterns of activity across presentations of different images. Figure~\ref{corr} shows the average correlation across repetitions of the same image compared to the average correlation across images. To do this analysis, the data from the 113 images were extracted from each ROI, which yielded an extracted dataset of 451 images (112 x 4 repetitions, and 1 x 3 repetitions). Participant CSI4 was not included in this analysis due to the relative insufficient amount of data because of early termination. A Pearson correlation was then calculated for each comparison across repetitions (e.g., correlate repetition 1 with repetition 2, correlation repetition 1 with repetition 3, etc.). The correlations across each pairwise repetition were then averaged for each image to provide a measure of similarity within image (i.e., across repetitions). The pairwise correlation for each image, for each repetition, was also made across all other images (from the pool of 451) and then averaged to give a measure of similarity across different images. The difference between these average correlations provides a measure of the reliability of the signal on a per trial basis.  

If there is reliability in the signal, the correlation across repetitions should be considerably higher than the correlation across images. This is indeed what we found.

\subsubsection*{Representational Similarity Analysis}

One type of analysis used in the literature thus far to compare computational models to brain activation patterns is representational similarity analysis (RSA). A popular comparison \cite{Cichy2016,Groen2018}, and one relevant to our mission, is comparing BOLD activity to the unit activity in a convolution neural net (CNN) aimed at recognizing objects – AlexNet \cite{Krizhevsky2012}. We have done the same analysis as a reference point to put in context with other related studies.  We implemented this analysis by comparing the similarity space derived from the pattern of BOLD activity across each voxel in a given region of interest to the similarity spaced derived from feature space of AlexNet for the set of scenes. We used 4,916 scenes in this analysis, where each scene is presented on a single trial, i.e., the trials in which a scene was repeated were not included in this analysis. For voxel space, the BOLD signal was extracted from each voxel of a given ROI such that we had a matrix of images by voxels (see Region of Interest Analysis). To measure similarity, the cosine distance of each pairwise comparison across images were made. This was performed for each ROI (N = 10, PPA, RSC, OPA, LOC, EarlyVis, in each hemisphere, see Region of Interest Analyses). In model space, we extracted the weights from each layer of AlexNet. All images were passed through an ImageNet pre-trained AlexNet, and all layer weights were extracted. Similarities were then computed for all weights in their original shape using the cosine distance metric across each pairwise comparison of images. This was then computed for each layer of AlexNet, such that 7 (5 convolutional, and 2 fully connected) feature spaces are derived from AlexNet. We then used a Pearson correlation to compare the voxel similarity space of a given ROI (i.e., cosine distances) to the feature similarity space of a given AlexNet layer. 

\begin{figure}[th!]
\centering
\includegraphics[width=\textwidth]{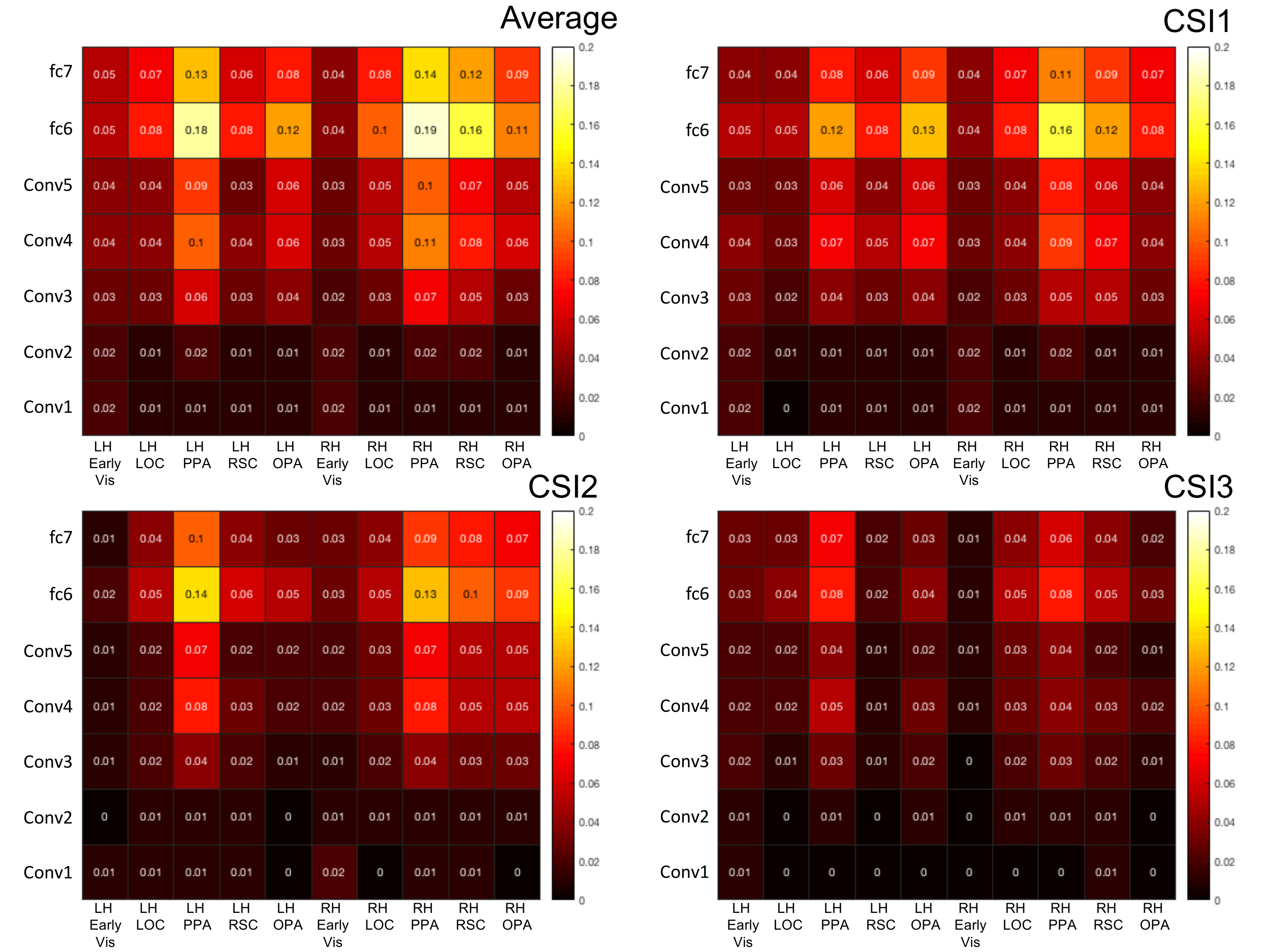}
\caption[RSA] {Representational similarity analysis heatmaps comparing the similarity of the 4,916 scenes in voxel space to the similarity of the 4,916 scenes in AlexNet features space. Comparisons were made across each ROI (columns) and each layer of AlexNet (rows).}
\label{RSA}
\end{figure}

 Figure~\ref{RSA} shows the RSA analysis for the average of three participants, as well as the results for each individual participant (the fourth was not included due to an incomplete set of data). The average was calculated by first averaging the cosine distance measurements and then correlating with AlexNet features. The results of this analysis demonstrate the typical pattern: high-level visual regions (e.g., PPA) correlate more strongly with higher layers of AlexNet (e.g., convolutional layer 5; presumably where high-level information, like semantics, is represented). Lower-level visual regions (e.g., EarlyVis) correlate stronger with lower layers (e.g., convolutional layer 2) compared to higher regions. Although we do not show a strong correlation with early visual regions and low levels of AlexNet, as commonly shown, we believe that this may result from our region of interest being across early visual regions as a whole and not confined to a retinotopically defined V1 or V2. However, the pattern of results in lower layers to brain regions (low level regions compared with high level regions) still holds. In addition, the correlation is at times lower than what has been reported in the literature. However, this may be a consequence of stimuli numbers, with low numbers inflating the correlation effect. Important to note from this analysis is the consistency of the results across hemispheres and across individual participants. 

\subsubsection*{ t-distributed Stochastic Neighbor Embedding Analysis}

A common type of analysis used in machine learning is t-distributed Stochastic Neighbor Embedding (t-SNE)\cite{vanDerMaaten2008}. The purpose of t-SNE is to embed high-dimensional data into a 2D space, such that high-dimensional data can be visualized. Importantly, t-SNE preserves similarity relations among high-dimensional data by modeling similar data with close points and divergent data with distant points. Similar to our RSA analysis, we perform t-SNE on the BOLD signal from each of the unique 4,916 (2,900 for CSI4) scenes trials. The BOLD signal was extracted from each voxel for each ROI for each participant. We visualize our t-SNE results with different categorical labels. Specifically, each t-SNE figure contains the \emph{same} data points/coordinates, and different labels are purely for visualization purposes. 

\begin{figure}[th!]
\centering
    \includegraphics[height= 0.89\textheight]{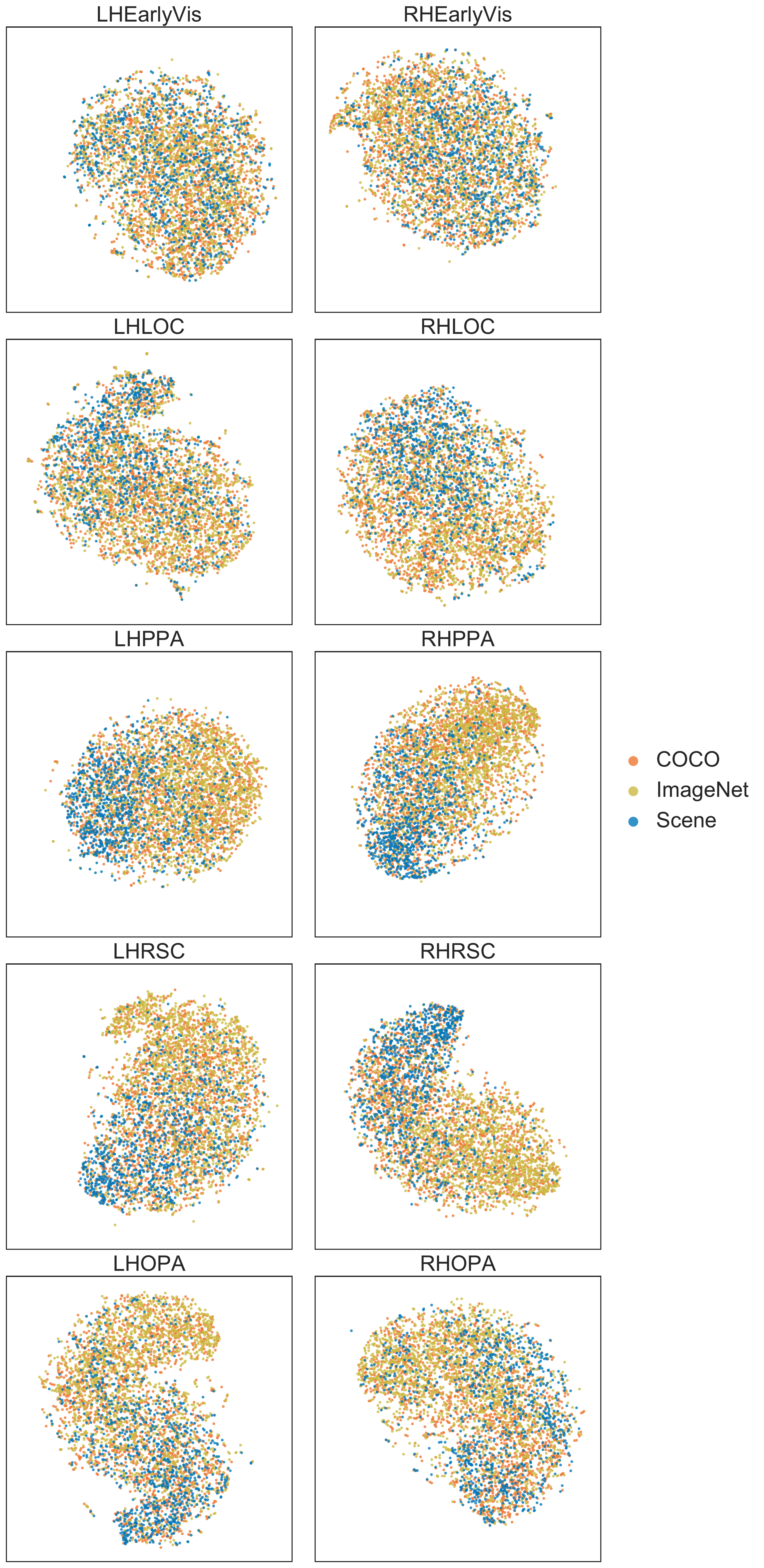}
    \caption[width=\textwidth]{t-SNE analysis on the 4,916 scenes in voxel space for CSI1. Image datasets (COCO, ImageNet, Scene) are used as labels.}
    \label{data_subj1}
\end{figure}

\begin{figure}[th!]
\centering
\includegraphics[height= 0.85\textheight]{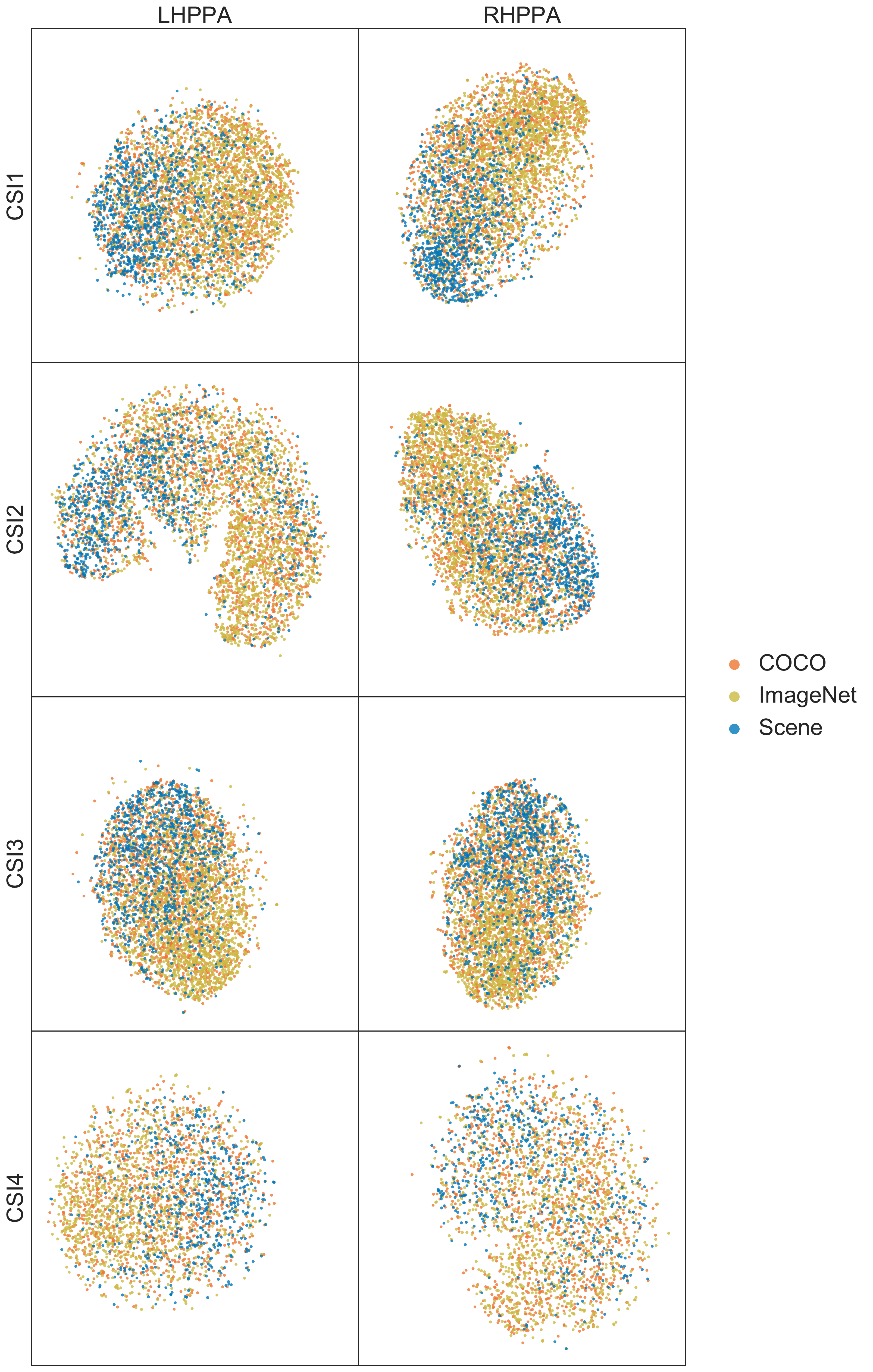}
\caption{t-SNE analysis on the 4,916 scenes in voxel space for all participants, in the PPA. Image datasets (COCO, ImageNet, Scene) are used as labels. Each row corresponds to an individual participant.}
\label{data_all}
\end{figure}

First, we examine the similarity space across the different image datasets. Specifically, we will be exploiting the implicit image attributes of these datasets: Scene contains whole scenes, ImageNet is focused on a single object, and COCO is in between with images of multiple objects in an interactive scene. Given the ROIs tend to process visual input with specific properties (e.g., category selectivity) we would expect to see a separation in t-SNE space of the different image datasets, especially with regard to Scenes vs. ImageNet. In Figure~\ref{data_subj1} the t-SNE results are visualized for CSI1 and each ROI. Here, the data points are labeled by the dataset the image belongs to (e.g. Scene, COCO, ImageNet). In ROIs commonly associated with scene processing (e.g. PPA, RSC, OPA), there were a clustering of the Scene images, a clustering of ImageNet images, and a more uniform scattering of COCO images. The strongest clustering of Scene images was observed in the PPA, regardless of participant (Figure ~\ref{data_all}). However, note that in lower-level visual regions (e.g. EarlyVis), we observed a uniform scatter for \emph{all} images, regardless of their datasets. This clustering and uniform scattering contrast reaffirms that higher-level scene regions have a stronger selectivity for processing categorical information. This pattern holds for all participants, and their t-SNE results for all ROIs can be seen on \url{http://BOLD5000.org}. 

\begin{figure}[th!]
\centering
\includegraphics[height= 0.85\textheight] {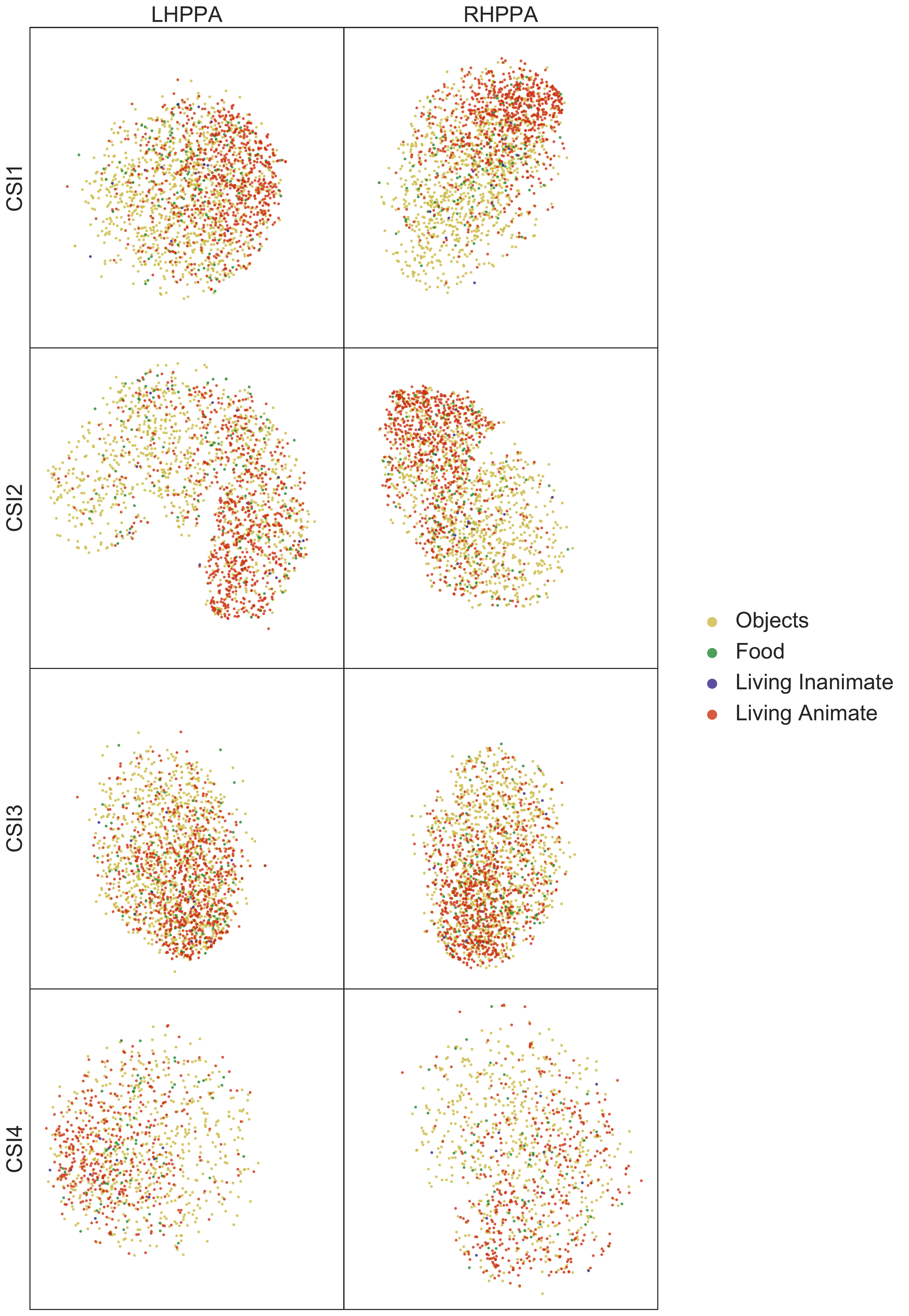}
\caption{t-SNE analysis on the 4,916 scenes (in voxel space for ImageNet images, for all participants, and in the PPA. ImageNet super categories (Objects, Food, Living Inanimate, Living Animate) are used as labels. Each row corresponds to an individual participant.}
\label{cat_all}
\end{figure}

Second, the t-SNE results are visualized with only ImageNet images, and our labels and categories are broken down further. Specifically, the images are labeled with ImageNet super categories. These super categories were created by using the WordNet hierarchy \cite{Miller1995}. For each image synset, which describes the image's object of interest, we found all of its WordNet ancestors. From 61 final WordNet categories, we labeled each one as ``Living Animate'',  ``Living Inanimate'',  ``Objects'', or  ``Food''. Images labeled as  ``Geography'' were removed because only 20 images applied. An example of our label mapping is  ``Dog'' to  ``Living Animate'' and  ``Vehicle'' to  ``Objects''. The t-SNE results are shown in Figure~\ref{cat_all}. Only the PPA region is presented to conserve space. The observations stated below also apply to other higher order regions, and all participants' ROIs for this t-SNE result are available on \url{http://BOLD5000.org}. In Figure~\ref{cat_all}, the data demonstrated that  ``Living Animate'' object based images clustered in a small area for all participants. However,  ``Objects''  based images are uniformly scattered in space. From the  ``Living Animate'' clustering, the results demonstrate that the higher order ROIs process images with living animate objects differently from objects, such as  ``cup''. From these results, it is evident that  ``Living Animate'' is a specific object category within the general object domain. Further, observing Figure~\ref{data_all} and Figure~\ref{cat_all}, one can see that the  ``Living Animate'' clustering occupies a separate spatial location than the Scene images clustering. Finally, it is important to emphasize that the results are consistent across hemispheres and across individual participants.

Using t-SNE, category selectivity in high-level visual regions was demonstrated using a data-driven method. The clustering observed demonstrates a small glimpse of the rich information inherent in the neural representations of each scene.

\newpage
\section*{Usage Notes}

The goal of this publicly available dataset is to enable joint research from various communities including neuroscience and computer vision. While this paper shows a glimpse of the richness in our data, we hope that our unique, large-scale, interdisciplinary dataset will provide new opportunities for more integrated analysis from various disciplines.

\section*{Discussion}
%----------------------------------------------------------------------------------------
%	SECTION 1
%----------------------------------------------------------------------------------------
Marr's \cite{Marr1982} nearly four-decades-old dream of a singular vision science --- the intertwined study of biological and computer vision --- is, at long last, being realized. Fueled by dramatic progress in neuroimaging methods and high-performing computer vision systems, new synergies are arising almost daily. However, connecting these two domains remains challenging. Here, we tackle one of the biggest obstacles for integrating across these two fields: data. In particular, neural datasets studying biological vision are typically lacking in: 1) {\em size}; 2) {\em diversity}; and 3) stimulus {\em overlap}; relative to extant computer vision datasets. We address these three concerns in BOLD5000 in which we have collected a large-scale, diverse fMRI dataset across 5,254 stimulus images. Critically, the human neuroimaging data available in BOLD5000 is: 1) significantly larger than prior slow event-related fMRI datasets by an order of magnitude; 2) extremely diverse in stimuli; 3) overlaps considerably with standard computer vision datasets. At the same time, BOLD5000 represents a significant dataset for the study of human vision in and of itself. As mentioned above, it is, by far, the largest slow event-related fMRI dataset using real-world images as stimuli. Moreover, given the diversity of content within these images and the fact that our fMRI data covers the whole brain, BOLD5000 may be sufficient to cover a wide range of high-level vision experiments (in the context of automatic visual processing during non-task-related, free viewing).

While we believe the scale of the BOLD5000 dataset is a major step forward in the study of vision across biological and computer vision, we should acknowledge its limitations. First and somewhat ironically, one salient limitation is the total number of stimulus images. Although 5,000 is significantly more than the number of images included in previous human neuroimaging studies, it is still relatively small as compared to either human visual experience across one's lifespan or the million of images used to train modern artificial vision systems. Given the practicalities of running the same individuals across multiple neuroimaging sessions, scaling up future datasets will necessitate collecting partially-overlapping data across participants and then applying methods for ``stitching'' data together \cite{Bishop2014,Bishop2018}. Second, another obvious limitation is that our dataset includes only four participants.\footnote{We encourage others in the field to add to the BOLD5000 dataset by running one or more participants using our exact experimental protocol and stimuli --- see the Code Availability section above.} Again, the practicalities of human experimental science come into play, necessarily limiting how many suitable participants we were able to identify and run. However, we would argue that the sort of in-depth, detailed functional data we collected at the individual level are as valuable as small amounts of data across many participants. Indeed, there have been recent arguments for exactly the sort of ``small-N'' design we have employed here \cite{Smith2018}. Ultimately we see a straightforward solution to these two limitations: expansion of our number of stimulus images by another order of magnitude, but with subsets of the stimuli run across many more participants. Such an undertaking is not for the experimentally faint-of-heart --- we suggest that the best approach for realizing this scale of human neuroscience would involve a large number of labs collaborating to run subsets of the overall study with coordinated data storage, distribution, and analysis.

\section*{Acknowledgements}

NC participated in stimulus selection, stimulus pre-processing, stimulus analysis, experimental design, collecting fMRI data, t-SNE data analysis, writing the manuscript, public distribution of the data, creating the website, and consulted on the remaining sections of the project. 

JAP developed and tested the MRI protocols, participated in experimental design, collecting fMRI data, the MRI processing pipeline, writing the manuscript, public distribution of the data, and consulted for the remaining sections of the project. 

AG helped conceive the original project and
and write the manuscript. AG consulted regarding stimulus selection, stimulus analysis, experimental design, and t-SNE data analysis.

MJT helped conceive the original project and write the manuscript. MJT consulted regarding stimulus selection, experimental design, data analysis, and public distribution.

EMA helped conceive the original project and participated in stimulus selection, experimental design, fMRI pre-processing data, general fMRI data analysis (GLM, ROIs), specific subsequent data analysis (design validation, data validation, representational similarity analysis), writing the manuscript, public distribution of the data and consulted on the remaining sections of the project.

We thank Scott Kurdilla for his patience as our MRI technologist throughout all data collection. We would also like to thank Austin Marcus for his assistance in various stages of this project, Jayanth Koushik for his assistance in AlexNet feature extractions, and Ana Van Gulick for her assistance with public data distribution and open science issues. 

This dataset was collected with the support of NSF Award BCS-1439237 to Elissa M. Aminoff and Michael J. Tarr, ONR MURI N000141612007 and Sloan, Okawa Fellowship to Abhinav Gupta, and NSF Award BSC-1640681 to Michael Tarr.

Finally, we thank our participants for their participation and patience, without them this dataset would not have been possible.

\section*{Competing Financial Interests}

The author(s) declare no competing financial interests.

\section*{Data Citations}
\label{sec:dataCite}
Brain, Object, Landscape Dataset \href{http://bold5000.org}{BOLD5000} (2018)

\end{document}